\documentclass{article}

\usepackage{url,hyperref,fullpage}
\usepackage{paralist}
\usepackage{amsmath}
\usepackage{mathtools}
\usepackage{amssymb}
\usepackage{cite}
\usepackage{amsmath,amssymb,amsfonts}
\usepackage{algorithmic}
\usepackage{graphicx}
\usepackage{textcomp}
\usepackage{xcolor}
\usepackage{color}

\usepackage{wrapfig}
\def\BibTeX{{\rm B\kern-.05em{\sc i\kern-.025em b}\kern-.08em
		T\kern-.1667em\lower.7ex\hbox{E}\kern-.125emX}}
\usepackage[linesnumbered,noresetcount,ruled,vlined]{algorithm2e}

\newcommand{\etal}{\emph{et al.}\xspace}
\newcommand{\eg}{\emph{e.g.,}\xspace}
\newcommand{\ie}{\emph{i.e.,}\xspace}
\newcommand{\Ie}{\emph{I.e.,}\xspace}
\newcommand{\sP}{\mathcal{P}}
\newcommand{\algSize}{smaller} 
\newcommand{\algSizeSmall}{smaller} 

\newcommand{\remove}[1]{}

\newtheorem{definition}{Definition}[section]
\newtheorem{theorem}{Theorem}[section]
\newcommand{\bigO}{\mathcal{O}}

\newenvironment{proofsketch}{\noindent\textbf{Proof.}}{\hfill$\blacksquare$}

\usepackage{booktabs}

\title{{A Self-stabilizing Control Plane for the Edge and Fog Ecosystems}\\ (preliminary version)}

\author{Zacharias Georgiou~\footnote{Department of Computer Science, University of Cyprus. Email: \{zgeorg03, chryssis, gpallis\}@cs.ucy.ac.cy} \and 
	Chryssis Georgiou~$^{\ast}$ \and
	George Pallis~$^{\ast}$ \and
	Elad M.\ Schiller~\footnote{Computer Science and Engineering, Chalmers University of Technology. Email: elad@chalmers.se} \and
	Demetris Trihinas~\footnote{Department of Computer Science, University of Nicosia. Email: trihinas.d@unic.ac.cy}
}

\begin{document}
	
	\maketitle
	
	\begin{abstract}
		Fog Computing is now emerging as the dominating paradigm bridging the compute and connectivity gap between sensing devices (a.k.a. ``things'') and latency-sensitive services. However, as fog deployments scale by accumulating numerous devices interconnected over highly dynamic and volatile network fabrics, the need for self-configuration and self-healing in the presence of failures is more evident now than ever. Using the prevailing methodology of self-stabilization, we propose a fault-tolerant framework for distributed control planes that enables fog services to cope and recover from a very broad fault model. Specifically, our model considers network uncertainties, packet drops, node fail-stop failures and 
		violations of the assumptions according to which the system was designed to operate, such as an arbitrary corruption of the system state. Our self-stabilizing algorithms guarantee automatic recovery within a constant number of communication rounds without the need for external (human) intervention. To showcase the framework's effectiveness, the correctness proof of the proposed self-stabilizing algorithmic process is accompanied by a comprehensive evaluation featuring an open and reproducible testbed utilizing real-world data from the intelligent transportation domain. Results show that our framework ensures a fog ecosystem recovery from faults in constant time, analytics are computed correctly, while the overhead to the system's control plane scales linearly towards the IoT load. 
	\end{abstract}

	\section{Introduction}
	Fog and Edge Computing are the technologies enabling computation at the network extremes, such as on downstream data, on behalf of cloud services, and upstream data, on behalf of IoT services~\cite{Shi2016}. The rationale of fog computing is that computing should happen at the proximity of the data source with the ``fog'' constituting any compute and network resources along the path between the data and the cloud. In this context, the ``edge'' differs from traditional sensing devices in that sensory data are processed in proximity and converted from raw signals to contextually relevant information~\cite{Trihinas2017}.

	In light of this, recent advancements in fog computing suggest using \textit{cloudlets} as intermediate compute platforms between IoT devices (edge devices) and the cloud which allow users to exploit the analytic power of the cloud without incurring the high latency in communicating with remote clouds~\cite{BITTENCOURT2018}. A cloudlet (also referred as a foglet, gateway, microcloud) can be a single server or a small cluster of co-located servers that form a (virtual) pool of shared resources but from an external viewpoint are considered a single entity~\cite{He2018}. Compared to traditional datacenters, a cloudlet features much more limited resources, albeit its proximity to IoT devices makes it appealing for offloading compute tasks and receiving timely responses.  
	
	Although fog computing brings the computation closer to delay-sensitive services, the challenges restricting the cloud paradigm still remain as the pace of generated data continues to rise~\cite{Trihinas2018}. Now, these overwhelming volumes of data not only have to be processed in time, but must be processed on, arguably, ``weaker'' hardware with potential nodes being vehicles, sensors, wifi access points, drones, cameras, and even wearable devices. Also, fog infrastructure usually operates in geo-distributed and less controlled environments, with many applications competing for limited resources against high-priority services (\eg 5G)~\cite{Trihinas2018b}. Consequently, failures due to hardware limitations and network uncertainties are highly likely at the fog continuum spanning between users, things, and clouds~\cite{Alarifi2019}.  To maintain high availability, fog infrastructure must be resilient to both node and network failures. Thus, self-managing and self-healing solutions are required for fog ecosystems. IoT services must be able to recover from any issues that arise during their lifetime. In this context, it is critical to ensure continuous operation and recoverability at scale even in the event of failure without human involvement. In particular, cloudlets must satisfy the increasingly stringent fault-tolerance specifications of today's internet-enabled systems. In the current fog computing paradigm, fault-tolerance must be implemented to both preserve the system state locally at the edge and ensure the accuracy of analytics computations,  especially in the case of a node failure or intermittent long-distance network connectivity problems. 

	We propose to address the challenge of dependable fog computing by using a fault-tolerant control plane that ensures service availability and data freshness in spite of the dynamic nature of the fog continuum. Via inter-connection of IoTs (edge devices), cloudlets and remote clouds, the proposed solution can tolerate network uncertainties, communication drops as well as cloudlet and IoT failures. In addition to these benign failures, our algorithms follow a very strong notion of fault-tolerance, called \emph{self-stabilization}~\cite{DBLP:books/mit/Dolev2000}, which has provided the Internet with automatic failure recovery as early as the 1980's~\cite{Perlman1999}. Self-stabilization ensures that the fog can recover after the occurrence of any temporary violations to the assumptions according to which the system was designed to operate. These violations can include, for example, state corruption, extreme number of node failures, network partitions or unexpected system reconfiguration. Once such transient violations occur, non-self-stabilizing systems cannot guarantee correct system behavior due to data loss or the propagation of corrupted information. The correctness proof of a self-stabilizing system is required to guarantee recovery, within finite time, after the occurrence of the last transient violation.
	
	\smallskip
	\noindent \textbf{Contribution and Research Outcome.~} This paper addresses the problem of how to tolerate and recover from run-time faults in distributed fog computing ecosystems. We consider a typical fog computing architecture, where edge devices are interconnected with remote clouds via network elements, denoted as cloudlets. Specifically: 
	\begin{itemize}[-]
		\item We introduce a self-stabilization framework for distributed control planes. The control plane is the core of the ecosystem and manages the network fabric with a global viewpoint and establishes the routing path of data serviced by geo-distributed cloudlets. To the best of our knowledge, we are the first to introduce a self-stabilizing framework for control planes enabled over fog and edge ecosystems~\cite{Hong2019}. 
		\item To deal with a broad fault model that includes both communication and node failures, our correctness proof details how the proposed self-stabilizing solution can  recover within  a  constant  number  of communication rounds, after the occurrence of transient faults, as required by~\cite{Dijkstra74}. 
		\item To illustrate both the effectiveness and low runtime footprint of our framework at scale, we introduce a thorough evaluation using real-world data and actual queries of interest from an intelligent transportation service. Our results are reproducible and the reference implementation (including configuration and test data) is open-source and available online\footnote{\url{https://github.com/UCY-LINC-LAB/Self-Stabilization-Edge-Simulator}}. Our experiments validate our analysis and show that even in the presence of severe failures, our solution can always recover in constant time while the network overhead scales linearly towards the IoT load.
	\end{itemize}

	\smallskip \noindent \textbf{Paper organization.~} Section~\ref{sec:related} reviews related research. Section~\ref{sec:system} presents the system model and objectives {before proposing the solution for realizing the system in Section~\ref{sec:alg}.} Section~\ref{sec:proof} presents the correctness proof. Section~\ref{sec:evaluation} presents the experimentation, followed by the conclusion.
	
	\begin{figure}[t!]
		\begin{\algSize}
			\centering
			\begin{tabular}{llll}
				\cline{2-3}
				\multicolumn{1}{l|}{}                    & \multicolumn{2}{l|}{~~~~~~~~~~~~~~~~~~~~~~~~~~~~~~~~~~~\textbf{Frequency}}                                                                               \\ \hline
				\multicolumn{1}{|l|}{\textbf{Duration}}  & \multicolumn{1}{l|}{\textit{Rare}}                          & \multicolumn{1}{l|}{\textit{Not rare}}                     \\ \hline
				\multicolumn{1}{|l|}{}                   & \multicolumn{1}{l|}{Any violation of the assumptions}         & \multicolumn{1}{l|}{Packet failures: omissions,}        \\
				\multicolumn{1}{|l|}{\textit{Transient}}          & \multicolumn{1}{l|}{according to which the system operates}           & \multicolumn{1}{l|}{duplications, reordering} \\
				\multicolumn{1}{|l|}{\textit{}}          & \multicolumn{1}{l|}{(as long as the code stays intact).}    & \multicolumn{1}{l|}{(assuming communication}      \\
				\multicolumn{1}{|l|}{\textit{}} & \multicolumn{1}{l|}{This can result in any state corruption.} & \multicolumn{1}{l|}{fairness holds).}                                   \\ 
				\hline
				\multicolumn{1}{|l|}{\textit{Permanent}} & \multicolumn{2}{l|}{~~~~~~~~~~~~~~~~~~~~~~~~~~Fail-stop failures.}                                                                                    \\ \hline
				\vspace*{0.25em}
			\end{tabular}
			\includegraphics[clip=true,scale=0.65]{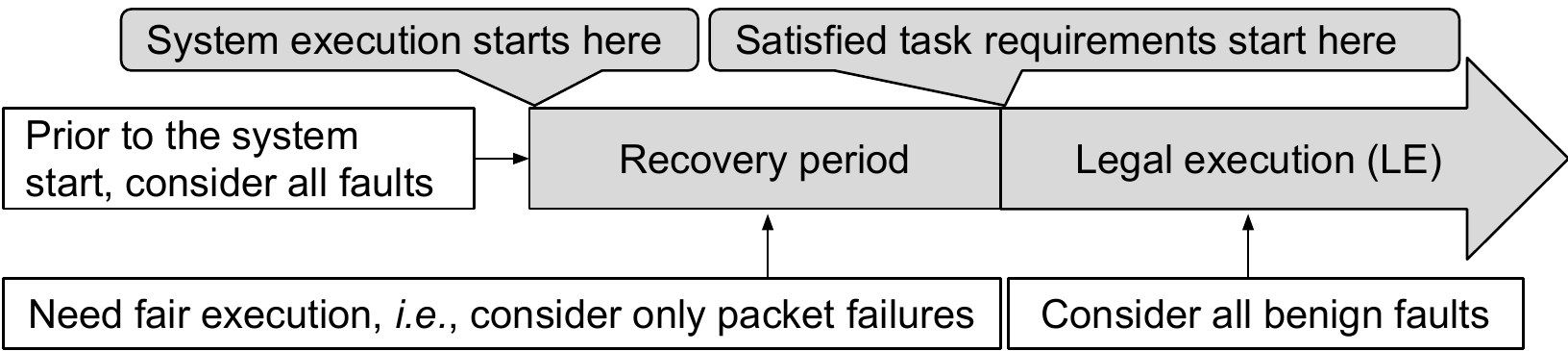}\\
			\caption{\label{fig:self-stab-SDN}\footnotesize{Table details the fault model and the chart illustrates when each fault set is relevant. The chart's gray shapes represent the system execution, and the white boxes specify the failures considered to be possible at different execution parts and recovery guarantees of the proposed self-stabilizing algorithm. The set of benign faults includes both packet failures and fail-stop failures.}}
		\end{\algSize}
	\end{figure}

	\section{Related Work}
	\label{sec:related}
	
	Fog and edge infrastructures are typically composed by hundreds of thousands to millions of heterogeneous and interacting components, which lead to the emergence of different types of faults. A major challenge in fog and edge computing is to define the fault and failure coverage required to provide high QoS~\cite{Harchol2018}. Faults may occur either simultaneously or in any aspect of system operations ranging from application to hardware, and may have several causes, including insufficient memory, performance interference, system utilization, network congestion, server faults, application crashes, etc. 
	Due to these challenges, existing work on fault-tolerance in large-scale distributed systems often have limitations in terms of practicality and performance guarantee. In~\cite{Harchol2018}, authors introduce CESSNA, a framework that provides consistency guarantees for stateful edge applications. CESSNA uses the Fault-Tolerant MiddleBox~\cite{Sherry2015}, which adopts the classical approach of ``rollback recovery'' where a system uses information logged during normal operation to correctly reconstruct state after a failure. In~\cite{Wang2019}, authors present a fault-tolerant messaging architecture for edge systems. The fault-tolerance is achieved by introducing timing bounds that capture the relation between service parameters and loss-tolerance 
	requirements. In~\cite{Wang2019T}, a fault-tolerant framework for data transmission in fog computing is introduced. The proposed fault-tolerance mechanism combines the advantages of Directed Diffusion and Limited Flooding to enhance the reliability of data transmission. We note that none of these solutions provides a holistic approach for addressing the fault-tolerance in edge and fog ecosystems. 
	
	Our framework fits naturally in distributed control planes, such as Istio and Linkerd~\cite{istio,linkerd}, that decouple operational control, policy enforcement and behavior telemetry from the business logic of distributed network fabrics and microservices. These frameworks provide fault-tolerance in the form of timeouts and (number of) retries for labelling nodes servicing HTTP requests as failed. In turn, circuit breaking is provided to safe-guard nodes overwhelmed by requests so that nodes ``fail fast'' when requests exceed the denoted limit.
	Thanks to our self-stabilizing algorithmic process, distributed control planes are introduced to a very strong notion of fault-tolerance on network uncertainties, communication drops, configuration  errors, arbitrary transient violations,  cloudlet  and  IoT  fail-stop  failures. In turn, no combination of faults can yield the system execution or corrupt data computations.
	
	In the context of self-stabilizing algorithms and IoT, Siegemund \etal~\cite{Siegemund2018} present a self-stabilizing publish/subscribe middleware for IoT applications. Their basic idea is that fault-tolerance is ensured through the construction of a distributed self-stabilizing data structure based on a virtual ring. However, operations over this ring take $\bigO(n)$ time even in the absence of failures, where $n$ is the ring size. Canini \etal~\cite{DBLP:conf/icdcs/CaniniSSSS18} present a self-stabilizing distributed control plane for software-defined networks (SDNs). Their work assumes that all nodes are either client hosts, switches or controllers. The algorithm stabilizes within $\bigO(d^2 n)$, where $d$ is the network diameter and $n$ is the number of nodes. Chattopadhyay \etal~\cite{Chattopadhyay2019} integrate an SDN control plane with the in-network processing infrastructure that can offload IoT services. They use a single centralized service deployment controller and lightweight SDN micro-controllers ($\mu C$). They {mention} that their algorithm for $\mu C$ placement is self-stabilized with a linear convergence time (but no formal proof is provided). We provide both analytical and empirical proof for convergence in constant time. The state-machine replication technique used in this paper is inspired by practically-self-stabilizing virtual synchrony~\cite{DBLP:journals/jcss/DolevGMS18}. However, the proposed self-stabilizing solution has a much easier to understand leader election mechanism than the one in~\cite{DBLP:journals/jcss/DolevGMS18}. Moreover, our self-stabilizing solution stabilizes in constant time whereas the one in~\cite{DBLP:journals/jcss/DolevGMS18} does not have a bounded stabilization time (by the definition of the solution criteria of practically-self-stabilizing systems).  
	
	While interesting and relevant, the above works do not address the impact of strong fault-tolerance in a hierarchical network organization that includes cloud infrastructure, cloudlets that are placed at the network edge and IoT devices. Our recovery time is within $\bigO(1)$ and our placement mechanism convergence is within $\bigO(1)$. We base our proofs on the definition of self-stabilizing systems~\cite{DBLP:books/mit/Dolev2000}. The definition requires the entire system to use bounded memory and recover after the occurrence of any transient violation of the assumptions according to which the system was design to operate. To the best of our knowledge, this is the first work that introduces a self-stabilizing control plane for the edge and fog ecosystems.



	\section{The System}
	\label{sec:system}
	{Informatics is a science of abstractions, and a main difficulty consists in providing users with a ``desired level of abstraction and generality --- one that is broad enough to encompass interesting new situations, yet specific enough to address the crucial issues''~\cite{DBLP:journals/dc/FischerM03}. This work provides a model that has the right-level of abstraction for the case edge computing since it allows both analytical and experimental study of the problem.}
	We consider a fog computing system comprised of sets of nodes, such as the one of cloudlets $C$ and IoT devices $S$, as well as a remote cloud infrastructure, which we refer to as the Cloud. Each cloudlet features specified communication, computation, and storage capabilities.  Each cloudlet is associated with a wireless access point covering a local area, referred to as a cell. The cloudlets in $C$ form a shared resource pool that can serve the system collaboratively, \eg aggregating IoT data and forwarding it to the Cloud. We assume that the cloudlets can share (over the Internet) such aggregated data with the Cloud by accessing a shared repository. The Cloud can use the repository to instruct cloudlets, \eg which queries the IoTs need to serve (edge devices), or provide advice the cloudlets on how to organize themselves, \eg propose the most-suitable leader according to the cloudlet specified capabilities and statistics gathered by the Cloud. The cloudlets themselves are intra-connected by backhaul links. We assume that, in the absence of failures, the quality of service of these links allow to send data and control messages in a timely manner --- this is in contrast to the communications between the cloudlets and the Cloud, which we assume to be asynchronous by nature. The control plane manages and configures the cloudlets to route traffic and enforce service placement with IoT devices.
	
	
	

	\smallskip \noindent \textbf{Objectives.~~}
	We aim at developing a fault-tolerant framework for distributed control planes that enables large-scale fog services to cope well with communication uncertainties and a broad fault model 
	without 
	service downtime or the need for external (human) intervention. Next, we discuss the development objectives of the proposed self-stabilizing solution before specifying the system requirements.
	
	\begin{itemize}[-]
		\item \textsf{O1.} The Cloud, cloudlets and IoTs (edge devices) should be able to exchange messages within a constant number of messages and communication rounds per information update.
		\item \textsf{O2.} The memory space and compute time of any system entity must always be bounded and network traffic scale linearly to the number of system entities.
		\item \textsf{O3.} The presence of a constant number of benign faults (Figure~\ref{fig:self-stab-SDN}) must not degrade the system performance beyond the bounds that are imposed by the system communication and processing delays. \Ie objective \textsf{O1} must not be violated in the presence of benign faults (and the absence of violations considered in objective \textsf{O4}).
		\item \textsf{O4.} We also consider arbitrary transient violations of the assumptions according to which the system was designed to operate (as long as the algorithm code stays intact). After the occurrence of these violations, the system must recover autonomously within a constant number of communication rounds and return to satisfy the task specifications. By autonomous we mean the absence of external intervention (of a human or a system component that is not part of the proposed framework).
	\end{itemize}
	%
	\begin{figure}
		\begin{flushleft}
			\centering
			\includegraphics[page=1,scale=0.475,trim = {0 0 0 1.5cm}]{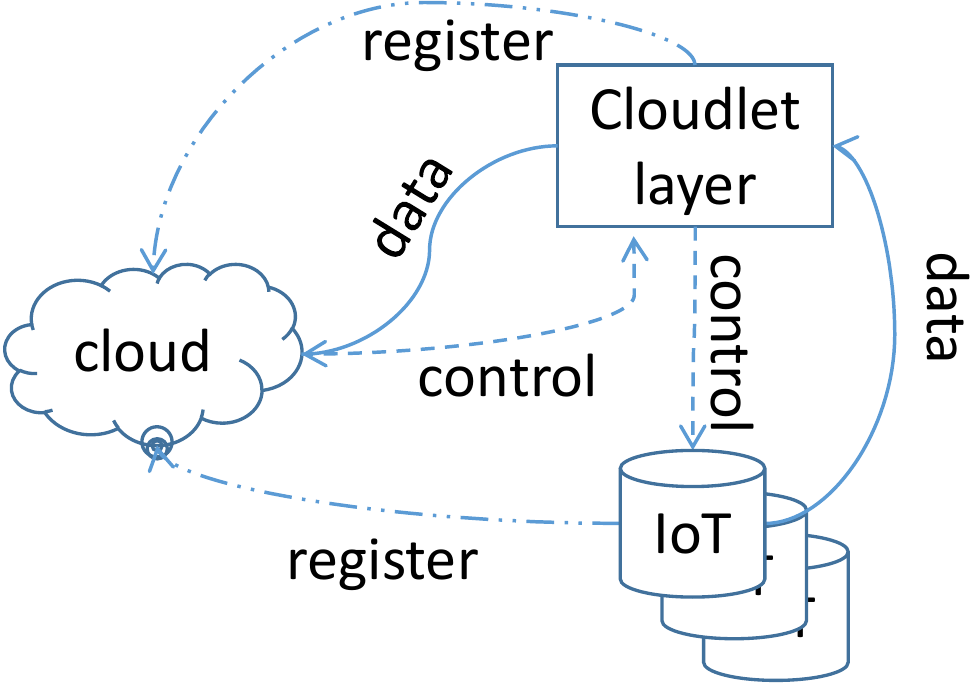}
		\end{flushleft}
		\caption{\label{fig:1st}System overview}
	\end{figure}
	
	\smallskip \noindent \textbf{Specifications.~~}
	The control plane for the edge organizes the \emph{cloudlet layer} (Figure~\ref{fig:1st}), such that in the presence of communication and node failures cannot disrupt the execution of services, such as IoT queries. In detail, we require the implementation of the following functionality: 
	
	(i) The cloudlet and IoT \emph{registration} allows the Cloud to include individual nodes in the system (Figure~\ref{fig:1st}). A node is allowed, after a predefined delay and local cleanups, to register again when it notices that it became disconnected from the system due to failures. Note that the latter case is rare, and thus, it should not repeatedly consume significant system resources. 
	
	\begin{figure}
		\begin{flushleft}
			\centering
			\includegraphics[page=2,scale=0.4375, trim = {0 0 0 1cm}]{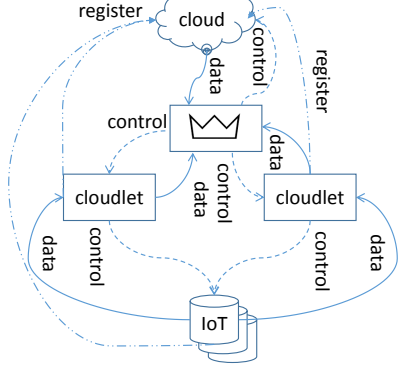}
		\end{flushleft}
		\caption{\label{fig:2nd}The leader-based architecture}
	\end{figure}
	
\begin{figure}[h!]
		\begin{flushleft}
			\centering
			\includegraphics[page=3,scale=0.4375,trim = {0 1cm 0 1.3cm}]{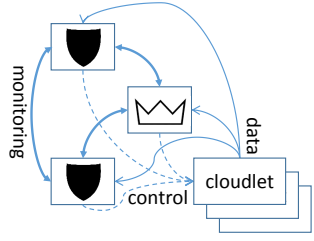}
		\end{flushleft}
		\caption{\label{fig:3rd}The inner-structure of the proposed cloudlet layer}
	\end{figure}

	(ii) The \emph{query} functionality allows the Cloud to request the flow of information according to a model that the IoTs (edge devices) are to update periodically. That is, given the Cloud's current belief about the query result, the specified IoTs (edge devices) will update the system whenever the collected sensory information deviates from the model. The cloudlet aim here is to aggregate these updates so that a concise query result arrives to the Cloud. Since this needs to be done in the presence of communication and node failures, each IoT should send its updates to a set of cloudlets and the latter should acknowledge (Figure~\ref{fig:2nd}). The cloudlets then should use a {\em leader} to unify their updates and forward concise query results to the Cloud. The cloudlet layer must function well in case of a failing leader. Therefore, a set of cloudlets, called {\em guards}, should monitor the leader's activity and guarantee query result delivery until the system decides on a new leader (Figure~\ref{fig:3rd}).              
	
	(iii) The management of \emph{general purpose services} can help to overcome capability differences among individual nodes via task load-balancing. Such tasks can be initiated by IoT users that need to leverage on the cloudlet capabilities. Also, cloud services may wish to avoid communication-intensive computations, such as virtual traffic light that base its decisions on the current road traffic conditions that different vehicles report. The fault-tolerant management of such services can be based on state-machine replication that is well-synchronized with query operations. 
	
	\setlength{\textfloatsep}{0pt}
	
	\LinesNumbered
	
	\begin{algorithm*}[t!]
		\begin{\algSize}		
			
			\BlankLine
			
			\noindent \textbf{Registers shared between the modules in algorithms~\ref{alg:Cloud},~\ref{alg:cloudlets}, and~\ref{alg:multVirSynShort}} 
			$\emph{\text{info}}$: has the form of $(devices$, $cloudlets$, $leader$, $guards)$, where the field $devices$ is a set of IoT devices, their models and the information needed for failure detection; $cloudlets$ is a set of cloudlets and the information needed for failure detection; $leader$ of the form $(seq,id)$ is the cloudlets' current leader and an associated sequence number; $guards$ is a set of cloudlets ids (a subset of $cloudlets$) that have been selected as guards;
			
			\BlankLine
			\tcc{~~~~~~~~~~~~~~~~\textbf{\textrm{the module for the self-stabilizing cloud (Algorithm~\ref{alg:Cloud})}}}
			\BlankLine

			\noindent \textbf{Local variables:} 
			$newCloudlet$/$newIot$: new cloudlets and IoTs (edge devices) and their models;
			$sequence$: leadership number;
			
			\BlankLine
			
			\noindent \textbf{do forever} \Begin{
				
				\If{the reset procedure is inactive and fresh information was recived from all trusted (not to be faulty) cloudlets}{
					
					Use $newIot$, $newCloudlet$ and fault detection information for updating $devices$ and $cloudlets$, respectively\; 
					\lIf{$leader \notin cloudlets$}{elect a leader with $sequence\text{++}$}
					\lIf{$sequence =\mathit{MAXINT}$}{invoke the reset procedure}
					\lIf{$guards \cap cloudlets = \emptyset$}{select new $guards$}
					
				}
				
				Once the reset procedure is done, initialize all local variables\;
			}

			\textbf{upon} registration request arrival from an IoT or a cloudlet, update the set $newIot$ and $newCloudlet$, respectively\; 
			
			\textbf{upon} RESET message arrival, invoke the distributed reset procedure\; 

			\BlankLine
			\tcc{~~~~~~~~~~~~~~~~\textbf{\textrm{the module for IoT (Algorithm~\ref{alg:IoT})}}}
			\BlankLine

			\noindent \textbf{Local variables for the IoT module:} 
			$model$: a data structure that encodes the recent sensory readings;
			$\mathit{cloudletModel}$: recent model received from the cloudlet;
			$cloudletList$: dissemination point list;
			$\mathit{lastUpdate}$: time of the last update reception from a cloudlet;
			$\mathit{msgseq}$: a positive integer used as a sequence number for messages sent to cloudlets; 
			$\mathit{MSG}$: a set that stores the highest message sequence received;
			
			\BlankLine
			
			\noindent \textbf{do forever} \Begin{
				
				\lIf{$\mathit{lastUpdate}$ was too long ago}{initialize and register at the Cloud this IoT}
				\ElseIf{an update is needed}{
					\lForEach{cloudlet in $cloudletList$}{send $\langle \mathit{msgseq}\text{++}, model \rangle$}
					\lIf{$\mathit{msgseq} =\mathit{MAXINT}$}{invoke the reset procedure}
				}
			}
			
			\noindent \textbf{upon} $m=\langle seq, list, model \rangle$ arrival from a cloudlet \textbf{do} \{\textbf{if} {$m$'s $seq$ is fresh} \textbf{then} {update $cloudletList$, $\mathit{cloudletModel}$, $lastUpdate$ and $\mathit{MSG}$} reply;\}
			
			\noindent \textbf{upon} reply $m=\langle seq \rangle$ arrival from a cloudlet, update $\mathit{msgseq}$\;

		\end{\algSize}
		\caption{A high-level overview on algorithms~\ref{alg:Cloud} and~\ref{alg:IoT}}
		\label{alg:wordCloudletsI}
	\end{algorithm*}
	
	\begin{algorithm*}[t!]
		\begin{\algSize}		
			
			\BlankLine
			\tcc{~~~~~~~~~~~~~~~~\textbf{\textrm{the module for the self-stabilizing cloudlet (Algorithm~\ref{alg:cloudlets})}}}
			\BlankLine

			\noindent \textbf{Local variables:} 
			$deviceSet$: a set of IoT devices and their most recently received models;
			$\emph{\text{agreegateInfo}}$: a set of data structures encoding aggregated sensory information;
			$msgc$: a positive integer used for ordering message sent to the leader and guards;
			$msgtoiot$: a positive integer used for ordering messages sent to IoT devices;
			$\mathit{MSGc}$: a set of $(id,seq)$ pairs that stores the highest message sequence received by cloudlet $id$;
			$\mathit{MSGSEQ}$: a set of $(id,seq)$ pairs that stores the highest message sequence received by IoT $id$;
			
			\BlankLine
			\noindent \textbf{do forever}
			\Begin{
				
				\lIf{the reset procedure is inactive and $i\notin cloudlets$}{initialize and register at the Cloud this cloudlet}
				
				\ElseIf{the reset procedure is inactive}{
					Use $devices$, $deviceSet$ and $cloudlets$	to update $deviceSet$, $\mathit{MSGSEQ}$ and $\mathit{MSGc}$, respectively\;
					
					\lForEach{IoT $j$ that this cloudlet is responsible for}{send info about $msgtoiot\text{++}$, the cloudlets that are responsible for this IoT and the model of this IoT}
					
					\lForEach{$j$ that is a guard or a leader}{send info about $msgc\text{++}$ and the aggregated data received from the IoT that this cloudlet is responsible for}
					
					\lIf{$\mathit{msgseq} =\mathit{MAXINT}$}{invoke the reset procedure}						
				}
			}
			
			\noindent \textbf{upon} $m=\langle seq,model\rangle$ arrival from an IoT \textbf{do} \{Acknowledge $m$ and update $deviceSet$ and $deviceSet$\}
			
			\noindent \textbf{upon} $m=\langle seq,aggregated\rangle$ arrival from a cloudlet \text{do} \{Acknowledge $m$ and update $\mathit{agreegateInfo}$ and $\mathit{MSGc}$\}
			
			\noindent \textbf{upon} $m=\langle seq \rangle$ arrival from an IoT \textbf{do} \{update $msgtoiot$\}
			
			\noindent \textbf{upon} $m=\langle seq \rangle$ arrival from a cloudlet \textbf{do} \{update $msgc$\}
			
			\BlankLine
			\tcc{~~~~~~~~~~~~~~~~\textbf{\textrm{the module for self-stabilizing replication for guards and leader (Algorithm~\ref{alg:multVirSynShort})}}}
			\BlankLine

			\noindent \textbf{Local variables:} $replicaState[]$: an array of the state machine's replica, where $rep[i]$ refers to the one that processor $p_i$ maintains, and $\mathit{replicaState}[j]$ refers to the last arriving message from $p_j$ containing $p_j$'s $\mathit{replicaState}[j]$. $myLeader$ stores the identifier of the local leader. The term $view$ refers to the set of replicas that the leader considers to be up and connected, \ie they can participate in the emulation of the state-machine. $FD$ stores the processors that the (local) failure detector considers as active\; 
			
			\BlankLine
			
			\noindent \textbf{ do forever} \Begin{
				\lIf{the Cloud propses $\mathit{leader}$ to be this replica but $myLeader$ does not or the view is not all trusted (not to be failing) guards and this replica}{propose a view with this replica as a leader as well as all trusted (not to be failing) guards as members}
				
				\lIf{$myLeader$ refers to this replica but the Cloud proposes another trusted guard}{update $myLeader$ to the proposed one}
				
				\lIf{this replica is the leader and all replicas have completed a communication round}{compute the new state of the automaton and update $\mathit{replicaState}$}
				\lElse{update $\mathit{replicaState}$ according to the one of the leader replica and send your input to the leader}
				
				\lIf{this replica is a guard that is not $\mathit{lLeader}$ but $\mathit{lLeader}$ is suspected (to be failing)}
				{reset $myLeader$ and update $\mathit{data}$ about the local state of this replica}
				\lElseIf{the $myLeader$ is well-defined (not reset)}{send this replica's state to $leader$}
				
				\lIf{this replica is the $\mathit{lLeader}$}{broadcast this replica's state to $lGuards\cap FD$}		
			}
			
			\noindent \textbf{upon} $m$ arrival from a guard or a leader \textbf{do} \{update $\mathit{replicaState}$ with $m$;\}
			
		\end{\algSize}
		\caption{A high-level overview on algorithms~\ref{alg:cloudlets} and~\ref{alg:multVirSynShort}}
		\label{alg:wordCloudletsII}
	\end{algorithm*}

	\section{Proposed Solution}
	\label{sec:alg}
	{Algorithms~\ref{alg:wordCloudletsI} and~\ref{alg:wordCloudletsII} provide} a high-level description of our solution, and the details appear in algorithms~\ref{alg:Cloud},~\ref{alg:IoT},~\ref{alg:cloudlets} and~\ref{alg:multVirSynShort}, which implement the proposed solution to the above task specifications by considering the code to be executed by the Cloud, IoT devices, cloudlets, and respectively, the emulators of the replicated state-machine. Algorithm~\ref{alg:Cloud} assumes the {availability} of a self-stabilizing cloud infrastructure, such as~\cite{DBLP:conf/services/BinunBDKMYCLW14}.
	
	\noindent{{\bf Overview.} The Cloud periodically monitors the system and keeps track of the Cloudlets and IoT devices that are up and running. Based on this information, and according to some mapping, each cloudlet is associated with a list of IoT devices. The IoT devices periodically send their data (e.g., sensory information) to their associated cloudlet(s). Instead of having each cloudlet to report directly to the cloud, each cloudlet reports its collected data to a leader. The leader is the one that collects and aggregates all data and reports it to the Cloud (via shared registers). The above constitutes a ``normal''(fault-free) operation. However, due to unexpected transient faults or more permanent faults (e.g., a cloudlet fail-stopping), as well as the need for bounded counters, additional checks must take place at the different components of the system. Algorithms~\ref{alg:Cloud},~\ref{alg:IoT}, and \ref{alg:cloudlets} present such details for the Cloud, the IoT devices and the Cloudlets, respectively. 
		Furthermore, in the event that the leader fail-stops, we do not want the data flow to the cloud to be suspended or critical information to be lost. To this respect, from the list of operational cloudlets, the Cloud also appoints a set of guards. The purpose of the guards is to monitor more frequently the status of the leader and in the event that the leader fail-stops, they report the latest collected data to the Cloud. Therefore, in Algorithm~\ref{alg:cloudlets},
		each cloudlet reports its collected data not only to the leader, but also to the guards. Since the leader and the guards need to maintain consistent information on the collected data (and on any other information the control plane could be maintaining), they run Algorithm~\ref{alg:multVirSynShort}, which realizes a self-stabilizing state-machine replication mechanism.
		In Section~\ref{sec:proof} we provide the correctness proof illustrating that our algorithmic framework can self-stabilize in a constant number of communication rounds, while Section~\ref{sec:evaluation} shows through a large testbed that there is no information loss even in the presence of multiple, different and randomly injected failures to the fog ecosystem.}
	
	We now proceed to present more details. We start by describing the registers that are shared by the nodes. Then, we go through the code according the above functionality list.
	
	
	\noindent {\bf\em Registers.~~}
	The shared register $\mathit{data}$ stores the aggregated sensory information that is collected by the IoTs (lines~\ref{ln:iotReg}--\ref{ln:msgseqSend}), aggregated by their corresponding cloudlets (line~\ref{ln:msgcSend}), and written by the leader (line~\ref{ln:writedata}).
	The Cloud and the cloudlet exchange control information via the shared registers $\mathit{info}$ and $\mathit{infoAck}$. The register $\mathit{info}$ includes the fields (IoT) $devices$, $cloudlets$, $leader$ and $guards$. 
	%
	%
	The register $\mathit{infoAck}$ is an array, such that the entry $\mathit{infoAck}[k]$ holds $p_k$'s acknowledgment, where $p_k \in C$ is a cloudlet and the acknowledgment includes all the fields of $\mathit{info}$. In detail, the Cloud, $p_{cloudID}$, stores its view on the system membership in $\mathit{info}$ (line~\ref{ln:botWrite}) and cloudlet $p_k$ acknowledges the reception of this information by copying the value of $\mathit{info}$ to $\mathit{infoAck}[k]$ (line~\ref{ln:readInfo}). Moreover, $p_{cloudID}$ selects, when needed, new cloudlets' leader (line~\ref{ln:newLeader}) and guards (line~\ref{ln:newGuards}).

	\noindent {\bf\em Registration.~~}
	IoTs (edge devices) and cloudlets register directly at the Cloud by sending a registration message (lines~\ref{ln:iotReg} and~\ref{ln:letReg}) after initializing their local variables and communication channels. This initialization guarantees that the joining node (or its communication channels) does not hold stale information. Once the registration message arrives to the Cloud, $p_{cloudID}$, the Cloud lists the joining node as a newcomer (lines~\ref{ln:newiot} and~\ref{ln:cloudlet}). These newcomers will be listed as the system's IoT devices and cloudlets (lines~\ref{ln:newIotEmpty} to~\ref{ln:newCloudletEmpty}) after the completion of the previous update round of these sets, which line~\ref{ln:botlInfot} assures. The proposed solution assumes access to unreliable failure detectors. This allows the Cloud not to wait for cloudlets that are suspected to be faulty as well as to remove failing nodes from the IoT and cloudlet sets.  
	
	\noindent {\bf\em Query.~~}
	We consider queries that are initiated by Cloud applications and require repeated updates. These queries include the Cloud current belief about the anticipated result, which we refer to as the query model. This allows IoT devices to reduce the number of times in which they transmit results to periodic queries since there is no need to transmit a result that fit the current belief of the Cloud according to the query model. 
	
	In detail, the registration procedure constructs up-to-date views on the sets of IoT $devices$ and $cloudlets$ in the shared register together with the current leader and guards. The proposed solution associates with each IoT the query description and model. This information is stored in $devices$. The cloudlets use a function, $myIoT()$, for mapping between them and the IoTs that they are responsible to communicate with (line~\ref{ln:msgtoiotSend}). 
	(A possible mapping could be to have the IoTs being assigned to the cloudlets in the same region, based on their proximity. Nevertheless, our system is independent on the specific mapping employed.)
	Cloudlets send the queries (along with their models) to these IoTs. The latter store the arriving information and acknowledge (lines~\ref{ln:iot:arrival} to~\ref{ln:msgseqRec}). Once in a predefined periodicity, the IoTs update the query results, if needed (line~\ref{ln:msgseqSend}). The cloudlets acknowledge the update arrival (lines~\ref{ln:msgseqRec} and~\ref{ln:MSGSEQack}). 
	The cloudlets in turn periodically aggregate the sensory information received by the IoTs and send it to the leader and the guards (line~\ref{ln:msgcSend}). The leader updates the shared repository with the query results (line~\ref{ln:writedata}), {whereas the guards serve as warm-backup leaders.} {We assume access to the functions $electLeader()$ and	$selectGuards()$ that for a given set of system cloudlets elect a leader and select guards, respectively.} 
	In electing a leader and guards, we may want nodes that are more stealth, maybe closer to the IoT devices or in the center of the coverage area (\eg in the center of the city); the leader/guard selection problem can be inherent to the fog service placement problem (FSPP)~\cite{DBLP:journals/soca/SkarlatNSBL17}, which is a different challenge in fog computing than the studied one. Nevertheless, in our system we could swap in/out FSPP algorithms and we are resilient to the algorithm in use.
	

	\noindent {\bf\em State-machine replication.~~}
	Since both the leader and the guards receive aggregated sensory information from the cloudlets, they need to be in sync with respect to this information. More generally speaking, the leader and the guards could provide additional service as part of the control plane. So, they need to coordinate their activities and maintain consistent state between them. The fact that the system is asynchronous, together with the need for self-stabilization, makes it quite a challenging task. To this respect, we have the leader and the guards to run Algorithm~\ref{alg:multVirSynShort}.
	
	
	
	The algorithm maintains a consistent state (aggregated sensory information) by performing multicast rounds coordinated by the leader. All necessary replica information (including the state) is maintained by each node in array $rep[]$ (line~\ref{VS:variables}), which is exchanged between the leader and the guards (lines~\ref{VS:send}--\ref{VS:receive}).  
	In detail, once a cloudlet realizes that it has become the leader (line~\ref{VS:newleader}), it proposes to install a {\em view} of the current members, which includes itself and the guards that according to its local failure detector have not fail-stopped. The guards start following the leader towards installing this view by adopting its proposal (line~\ref{VS:flrPropose}). Once the leader sees that 
	the view members have adopted its proposal (lines~\ref{VS:crdConditions} and \ref{VSm:switch}),
	it builds the new state based on the collected messages and states (lines~\ref{VS:crdPropose} and \ref{VS:updatestate})
	and proceeds to install the view. The guards adopt the leader's $rep$ -- including the (new) state (lines~\ref{VS:flrInstall} and \ref{VSm:adoptrep})
	completing in this way the installation of the view (lines~\ref{VS:crdInstall} and \ref{VS:installsview}). 
	The multicast rounds can now begin, which are coordinated
	by the leader (lines~\ref{VS:crdMulticast} and \ref{VSm:procedure}--\ref{VSm:rndIncr}) and
	followed by the guards (lines~\ref{VS:flrMulticast} and
	\ref{VSm:fetchFol}).
	The access to the application's message queue (commands to be executed by the
	state machine) is done via $\mathit{fetch}()$, which returns the next
	multicast message; the state transition function $apply(state,msg)$ applies the aggregated input array $msg$ to the replica's $state$ and produces the local side effects. Simply put, in our case,
	the input to the state machine is the aggregated sensory information, which is sent by the cloudlets to the leader and the guards in Algorithm~\ref{alg:cloudlets} (line~\ref{ln:msgcSend}) and stored by the latter in $agrregateinfo$ (line~\ref{ln:agrinfo}). So, essentially the
	multicast rounds of the state machine keep this information consistent among the leader and the guards. At the end of each multicast round, the leader updates the sensory information maintained in the shared register $data$ (line~\ref{ln:writedata}). 
	
	In the event of a leader fail-stop, and until the Cloud assigns a new leader (line~\ref{ln:newLeader}),
	the guards update the $\mathit{data}$ repository (lines~\ref{ln:leaderfail}--\ref{ln:guardwrite}), instead; this ensures a continual update of the sensory information (which, depending on the application, could be crucial). If there is a change in the set of guards (either due to a fail-stop or due to an update of this set by the Cloud), then the leader begins the procedure to install a new view (line~\ref{VS:newviewsl}) with the new membership, without
	the need of any external intervention (including that of the Cloud). 
	{The failure detector abstraction (defined in line~\ref{VS:variables}) can be implemented using heartbeats and counter thresholds (see for example~\cite{DBLP:conf/netys/BlanchardDBD14}), or using ``hello'' messages and timeouts in a more time-informed setting (as we do in our simulation study in Section~\ref{sec:evaluation}).}
	%
	
	\smallskip \noindent \textbf{\emph{Recovering the system state via global reset.~~}}
	Self-stabilization requires bounded space, which includes bounded counters. Counters can grow up to a predefined size $\mathit{MAXINT}$, \eg $2^{64}-1$. Under normal operation, and if say, a counter is incremented every nano-second, then this limit could be reached in approx. 146 years. However, a transient violation of the assumptions according to which the system was designed to operate can corrupt the counter and cause it reach $\mathit{MAXINT}$. In such a case (lines \ref{in:msgseqMaxed}, \ref{ln:countersMaxed}, and \ref{ln:cntMaxed}), the cloudlet or IoT {holding} this counter will send a \textsf{RESET} message to the Cloud, calling for a {global} system reset. The Cloud, upon receiving such a message (line~\ref{ln:reset}) or the $sequence$ counter reaches $\mathit{MAXINT}$ (line~\ref{ln:seqMaxed}),
	initiates the \emph{reset procedure}: it sets the shared register $\mathit{info}$ into $\bot$ (line~\ref{ln:seqMaxed} or line~\ref{ln:reset}),
	and waits until all non-faulty cloudlets have acknowledged this (via the shared array $\mathit{infoAck}$,  line~\ref{ln:empWrite}), before it unregisters all cloudlets and IoT devices (by setting $\mathit{info}$ into $(\emptyset,\emptyset,\emptyset,\emptyset)$) and flashes all its local variables. This causes each cloudlet (line~\ref{ln:letReg}) to register again after a local reset of the node state and its communication channels, following the registration procedure described above.
	Since the IoTs are no longer in $\mathit{info.devices}$, no cloudlet will contact them, causing each (non-faulty) IoT to timeout and hence also register again after a similar initialization procedure (line~\ref{ln:iotreg}).
	
	\begin{algorithm}[t!]
		\begin{\algSize}
			
			\noindent \textbf{Variables:} 
			$newCloudlet$/$newIot$: new cloudlets and IoTs and their models (bounded by $cloudletSetSize$);
			$sequence$: leadership number;
			
			\BlankLine
			
			\noindent \textbf{Shared registers:} 
			$\mathit{data}$: is a data structure that stores the sensory information, to be processed by the cloud depending on the application; it includes records of the form $(id,leader,round,dat)$, where $id$ is the cloudlet's unique id that included the context $dat$ in the data structure, at round $round$ of the state machine with leader $leader$; 
			$\emph{\text{info}}$: has the form of $(devices$, $cloudlets$, $leader$, $guards)$, where the field $devices$ is a set (bounded by $deviceSetSize$) of IoT devices, their models and the information needed for failure detection; $cloudlets$ is a set (bounded by $cloudletSetSize$) of cloudlets and the information needed for failure detection; $leader$ of the form $(seq,id)$ is the cloudlets' current leader and an associated sequence number; $guards$ is a set of cloudlets ids (a subset of $cloudlets$) that have been selected as guards;
			$\emph{\text{infoAck}}[cloudletSetSize]$: an array that stores the latest value of $\mathit{info}$ that each cloudlet has read;

			\BlankLine

			\noindent \textbf{Interface:} 
			$suspectedIot(set)$ and $suspectedCloudlet(set)$: return the sets of suspected to be faulty IoT devices and cloudlets, respectively; 
			$electLeader(set)$: returns the elected leader from $set$;
			$selectGuards(set)$: returns the set of guards from $set$;
			
			\BlankLine
			

			
			\noindent \textbf{do forever} \texttt{/* use predefined periodicity */} \Begin{
				
				{\textbf{let} $\mathit{lInfo}:=(lDevices,lCloudlets,\mathit{lLeader},lGuards)	:=\mathbf{read}(\mathit{info})$}\; 
				
				{\textbf{let} $\mathit{lInfoAck}	:=\mathbf{read}(\mathit{infoAck})$}\;

				\If{$\bot \neq \mathit{lInfo}  \land ( \{\mathit{lInfo} \} = \{ \mathit{lInfoAck}[k]	: k \in C\setminus suspectedCloudlet(C) \})$\label{ln:botlInfot}}{
					%
					$(lDevices,newIot) \gets ((lDevices \setminus \{ (k,\bullet) : k \in suspectedIot(lDevices)\}) \cup newIot,\emptyset)$\label{ln:newIotEmpty}\;
					
					$(lCloudlets,newCloudlet) \gets ((lCloudlets \setminus \{ (k,\bullet) : k \in suspectedCloudlet(lCloudlets)\}) \cup newCloudlet,\emptyset)$\label{ln:newCloudletEmpty}\;
					
					\lIf{$\mathit{lLeader}.id \not\in lCloudlets$}{$\mathit{lLeader} \gets (sequence\text{++}, electLeader(lCloudlets))$\label{ln:newLeader}}
					{\lIf{$sequence =\mathit{MAXINT}$}{\textbf{write}$(\mathit{info},\bot)$}\label{ln:seqMaxed}}
					\lIf{$(lGuards \cap lCloudlets) = \emptyset$}{$lGuards \gets selectGuards(lCloudlets\setminus\{\mathit{lLeader}.id\})$\label{ln:newGuards}}
					
					\textbf{write}$(\mathit{info},(lDevices,lCloudlets,\mathit{lLeader},lGuards))$\label{ln:writeInfo}\;
				}
				\lElseIf{$\{\bot,(\emptyset,\emptyset,\emptyset,\emptyset) \} \supseteq ( \{\mathit{lInfo} \} \cup \{ \mathit{lInfoAck}[k]	: k \in C\setminus suspectedCloudlet(C) \})$}{\textbf{write}$(\mathit{info},(\emptyset,\emptyset,\emptyset,\emptyset))$; $(newCloudlet,newIot,sequence)\gets (\emptyset,\emptyset,0)$}\label{ln:empWrite}
				\lElseIf{$\{\bot \} \subset ( \{\mathit{lInfo} \} \cup \{ \mathit{lInfoAck}[k]	: k \in C\setminus suspectedCloudlet(C) \})$}{\textbf{write}$(\mathit{info},\bot)$\label{ln:botWrite}}}
			
			\noindent \textbf{upon message} {$m=\langle \textsf{REGISTER}\rangle$} \textbf{arrival from IoT} $j$  \textbf{at time} $t$ \textbf{do} $newIot \gets (newIot \cup \{ (j,t,\bot) \})$\label{ln:newiot}\;
			
			\noindent \textbf{upon message} {$m=\langle \textsf{REGISTER}\rangle$} \textbf{arrival from cloudlet} $z$  \textbf{at time} $t$ \textbf{do} $newCloudlet \gets (newCloudlet~\cup \{ (z,t) \})$\label{ln:cloudlet}\;
			
			{\noindent \textbf{upon message} $m=\langle \textsf{RESET}\rangle$ \textbf{arrival from device} $k$  \textbf{do} 
				\textbf{write}$(\mathit{info},\bot)$\label{ln:reset}\;
			}
			\BlankLine

			\caption{Code for the self-stabilizing cloud $p_{cloudID}$.}
			\label{alg:Cloud}
		\end{\algSize}	
	\end{algorithm}

	\begin{algorithm}[t!]
		\begin{\algSize}
			
			%
			\noindent \textbf{Local state:} 
			$model$: a data structure that encodes the recent sensory readings;
			$\mathit{cloudletModel}$: recent model received from the cloudlet;
			$cloudletList$: a list (bounded by $cloudletListSize$) of dissemination points (ordered by descending priority);
			$lastUpdate$: time of the last update reception from a cloudlet (according to IoT's local time);
			$\mathit{msgseq}$: a positive integer used as a sequence number for messages sent to cloudlets; 
			$\mathit{MSG}$: a set of $(id,seq)$ pairs that stores the highest message sequence received
			by cloudlet $id$;
			
			\BlankLine
			\noindent \textbf{Interface:} 
			$update()$: receives the last sent $model$ and received $\mathit{cloudletModel}$ as well as the time in which that reception occurred ($lastUpdate$). The function then updates $model$ (and returns true) if the cloudlet model requires an update due to change in sensory input, a timeout due to a missing acknowledgment from the cloudlet or a change in the cloudlet model specifications; 
			
			\BlankLine
			
			\noindent \textbf{Function:} $iotInit()$: the IoT device first resets all variables dealing with Cloudlet data and control information as well as local data and control variables. Then it sends a special message $\mathit{INIT}$ to the Cloud, so that the Cloud removes all information about this device from the Cloudlets. Once this is done, the Cloud returns
			an acknowledgment to the device, and the function returns. 
			
			\BlankLine
			
			\noindent \textbf{do forever} \texttt{/* use predefined periodicity */} \Begin{

				\lIf{$(clock()-\mathit{lastUpdate})>\mathit{LIMIT}$}{$IoTinit();$ $\mathbf{send}(cloudID, \langle \textsf{REGISTER} \rangle)$\label{ln:iotreg}}
				\ElseIf{{$update(model,\mathit{cloudletModel},lastUpdate)$\label{ln:iotReg}}}{
					\lForEach{$id \in cloudletList$}{$\mathbf{send}(id, \langle \mathit{msgseq}, model \rangle)$\label{ln:msgseqSend}}
					$msgseq\gets msgseq + 1$ \texttt{/* if a message was sent */} 
					
					\lIf{$msgseq =\mathit{MAXINT}$}{$\mathbf{send}(cloudID, \langle \textsf{RESET} \rangle)$}\label{in:msgseqMaxed}
			}}
			
			\noindent \textbf{upon} $m=\langle seq, list, model \rangle$ \textbf{arrival from cloudlet} $j$ \textbf{at time} $t=clock()$  
			\Begin{
				\If{$m.seq > \mathit{MSG}|j.seq$}{$(\langle cloudletList,\mathit{cloudletModel}\rangle, lastUpdate) \gets (\langle m.list, m.model \rangle,t)$\label{ln:iot:arrival}\;
					$\mathit{MSG}\gets (\mathit{MSG} \setminus\{(k,\bullet) : k\notin cloudletList\ \vee k=j\})\cup (j,m.seq)$\;}
				$\mathbf{send}(j, \langle \mathit{MSG}|j.seq  \rangle)$\label{ln:modelAriivalAck}\;
				
			}
			\noindent \textbf{upon message} $m=\langle seq \rangle$ \textbf{arrival from} cloudlet $z$  \textbf{do}
			$\mathit{msgseq} \gets \max\{m.seq,\mathit{msgseq} \}$\label{ln:msgseqRec}\;

			
			\caption{Code for IoT $iot_i$}
			\label{alg:IoT}
		\end{\algSize}	
	\end{algorithm}
	
	\begin{algorithm}[t!]
		\begin{\algSize}
			
			\BlankLine
			
			\noindent \textbf{Local state:} 
			$deviceSet$: a set (bounded by $deviceSetSize$) of IoT devices and their most recently received models\;
			
			$\emph{\text{agreegateInfo}}$: a set of data structures encoding aggregated sensory information\;
			
			$msgc$: a positive integer used for ordering message sent to the leader and guards\;
			$msgtoiot$: a positive integer used for ordering messages sent to IoT devices\;
			$\mathit{MSGc}$: a set of $(id,seq)$ pairs that stores the highest message sequence received
			by cloudlet $id$\;
			$\mathit{MSGSEQ}$: a set of $(id,seq)$ pairs that stores the highest message sequence received
			by IoT $id$\;
			
			\BlankLine	
			\noindent \textbf{Shared registers:} 
			$\mathit{info}$ and $\emph{\text{infoAck}}$: as in Algorithm~\ref{alg:Cloud}\;

			\BlankLine
			
			\noindent \textbf{Interface:} 
			$aggregate(deviceSet)$: returns the aggregated sensory information\;
			
			$cloudletList(k,set)$: for a given IoT device $iot_k$ and a $set$ of cloudlets, this function returns the cloudlet list that $iot_k$ should use (prioritized in an descending order)\;
			
			$myIoT()$: projection of the IoTs that are within the cloudlet's responsibility\; 
			$cloudID$: the address of the Cloud\;

			\BlankLine
			
			\noindent \textbf{Function:} $cloudletInit()$: the cloudlet first resets all variables dealing with the data and control information of cloudlets and IoT devices as well as its local data and control variables. Then it broadcasts a special message $\mathit{INIT}$ to all other cloudlets, and to the Cloud so that the other cloudlets remove all  information about this cloudlet; the Cloud removes all relevant information about this
			cloudlet from the IoT devices. Once the cloudlet receives acknowledgments from all
			the cloudlets and the Cloud, the function returns.

			\BlankLine
			\noindent \textbf{do forever} \texttt{/* use predefined periodicity */} 
			\Begin{
				
				\textbf{let} $\mathit{lInfo}:=(lDevices,lCloudlets,\mathit{lLeader},lGuards)	:=\mathbf{read}(\mathit{info})$;  $\mathbf{write}(\mathit{infoAck}[i],\mathit{lInfo})$\label{ln:readInfo}\; 
				\lIf{$\mathit{lInfo}\neq \bot \land i\notin lCloudlets$}{\{$cloudletInit();$ $\mathbf{send}(cloudID, \langle \textsf{REGISTER} \rangle)$\label{ln:letReg})\}}
				
				\ElseIf{$\mathit{lInfo}\neq \bot$}{
					\lIf{$i\notin (lGuards\cup \{\mathit{lLeader}.id\})$}{$(\emph{\text{agreegateInfo}},\mathit{MSGc})\gets (\emptyset,\emptyset)$}
					$deviceSet \gets (deviceSet \setminus \{ (k,\bullet) : k \notin lDevices\})$\label{ln:deviceSetUpdate}\;
					$\mathit{MSGSEQ}\gets (\mathit{MSGSEQ}\setminus \{(k,\bullet) : k \notin deviceSet\})$\;
					$\mathit{MSGc}\gets (\mathit{MSGc}\setminus \{(k,\bullet) : k \notin lCloudlets\})$\;
					\textbf{let} $(iotAdd, msgAdd):=(0,0)$\;
					\lForEach{$(j, m) \in myIoT(lDevices,lCloudlets)$}{\{$\mathbf{send}(j, \langle msgtoiot, cloudletList(j,lCloudlets), m \rangle)$; $iotAdd \gets 1$\label{ln:msgtoiotSend}\}}
					
					\lForEach{$j \in lGuards\cup \{\mathit{lLeader}.id\}$}{\{$\mathbf{send}(j,\langle msgc,aggregate()\rangle)$; $msgAdd \gets 1$\label{ln:msgcSend}\}}
					$(msgtoiot,msgc)\gets (msgtoiot +iotAdd, msgc +msgAdd)$\;

					\lIf{$\text{MAXINT} \in \{msgc, msgtoiot\}$}{$\mathbf{send}(cloudID, \langle \textsf{RESET} \rangle)$\label{ln:countersMaxed}}

			}}

			\noindent \textbf{upon message} $m=\langle seq,model\rangle$ \textbf{arrival from} IoT $j$  \textbf{at time} $t$ \Begin{ 
				\If{$m.seq > \mathit{MSGSEQ}|j.seq$}{
					$deviceSet \gets (deviceSet \setminus \{(j, \bullet)\}) \cup \{(j, t, m)\}$\label{ln:deviceSetArrival}\;
					$\mathit{MSGSEQ}\gets (\mathit{MSGSEQ} \setminus\{(j,\bullet)\})\cup (j,m.seq)$\;}
				$\mathbf{send}(j, \langle \mathit{MSGSEQ}|j.seq  \rangle)$\label{ln:MSGSEQack}\;
			}
			
			\noindent \textbf{upon message} $m=\langle seq,aggregated\rangle$ \textbf{arrival from} cloudlet $z$  \textbf{at time} $t$ \Begin{ 
				
				\If{$i\in lGuards\cup \{\mathit{lLeader}.id\}$ $\wedge$ $m.seq > \mathit{MSGc}|z.seq$ }{$\emph{\text{agreegateInfo}} \gets \label{ln:agrinfo} (\emph{\text{agreegateInfo}} \setminus \{(z, \bullet)\}) \cup \{(z, t, m)\}$\label{ln:agre}\;
					$\mathit{MSGc}\gets (\mathit{MSGc} \setminus\{(z,\bullet)\})\cup (z,m.seq)$\;}  
				$\mathbf{send}(z, \langle  \mathit{MSGc}|z.seq \rangle)$\;
			}
			
			\noindent \textbf{upon message} $m=\langle seq \rangle$ \textbf{arrival from} IoT $k$  \textbf{do} {$msgtoiot \gets \max\{m.seq,msgtoiot \}$\label{ln:msgtoiotRec}}
			\BlankLine
			
			\noindent \textbf{upon message} $m=\langle seq \rangle$ \textbf{arrival from} cloudlet $z$  \textbf{do} {	$msgc \gets \max\{m.seq,msgc \}$\label{ln:msgcRec}}
			\BlankLine

			\caption{Code for cloudlet $p_i$}
			\label{alg:cloudlets}
		\end{\algSize}	
	\end{algorithm}
	
	\begin{algorithm}[!h]
		\caption{\label{alg:multVirSynShort}Self-stabilizing replication for guards and leader, code for cloudlet $p_i$}
		\begin{\algSizeSmall}
			\noindent \textbf{Interfaces:}
			$\mathit{fetch}()$ next multicast message, $apply(state, msg)$ applies the step $msg$ to $state$ (while producing side effects), $synchState(replica)$ returns a replica consolidated state, $synchMsgs(replica)$ returns a consolidated array of last delivered messages, $failureDetector()$ returns a vector of processor ids,
			$cloudID$ returns the address of the Cloud\;
			
			\BlankLine
			
			\noindent \textbf{Variables:}\label{VS:variables} $rep[]=\langle view\!=\!\langle ID, set \rangle, status\!\in\!\{\textsf{Propose},\textsf{Install}$, $\textsf{Multicast}\},$ $(multicast$ $round$ $number)$ $rnd$, $(replica)$ $state$, $(last$ $delivered$ $messages)$ $msg[n]$ $(to$ $the$ $state$ $machine)$, $(last$ $fetched)$ $input$ $(to$ $the$ $state$ $machine)$, $propV$ $=\langle ID$, $set \rangle$, $(recently$ $live$ $and$ $connected$ $component)$ $FD \rangle$: an array of the state machine's replica, where $rep[i]$ refers to the one that processor $p_i$ maintains, and $rep[j]$ refers to the last arriving message from  $p_j$ containing $p_j$'s $rep[j]$.
			$FD$ stores the $failureDetector()$ output, \ie the set of processors that the failure detector considers as active. 
			$myLeader$ stores the id of the local leader; $\bot$ if none. {The $view.ID$ (and $propV.ID$) is composed by the id and leader sequence installing the view, and counter $cnt$, in case the same leader installs a new view}\;

			\BlankLine
			
			\noindent \textbf{Shared registers:} 
			$\mathit{\mathit{info}}$ and $\mathit{\mathit{data}}$: as in Algorithm~\ref{alg:Cloud}\;
			
			\BlankLine
			
			\noindent \textbf{Macros:} \label{VSm:macros}
			%
			$roundProceedReady()= \{(\forall p_j \in view.set$: $rep[j].(view$, $status, rnd) = (view, status, rnd)) \lor ((status\neq\textsf{Multicast})$ $\wedge$ $[(\forall p_j$ $\in propV.set: rep[j].(propV,$ $status)= (propV,$ $\textsf{Propose}))$ $\lor$ $(\forall p_j \in propV.set:$ $rep[j].(propV, status)= (propV,\textsf{Install}))]\}$\label{VSm:switch}\; 
			
			$coordinatePropose()=\{(state,msg,status)\gets(synchState(rep),synchMsgs(rep),\textsf{install}\}$\label{VS:updatestate}\; 
			
			$coordinateInstall()=\{(view,status,rnd)\gets (propV,\textsf{Multicast},0)\}$\label{VS:installsview}\;
			
			$roundReadyToFollow()=\{rep[myLeader].rnd=0\lor rnd<rep[myLeader].rnd\lor rep[myLeader].(view \neq propV)\}$\label{VS:readytofollow}\; 
			
			$followPropose()= \{(status,propV)\!\!\gets\!\!rep[myLeader](status,propV)\}$\label{VSm:adoptProp}\; 
			
			$followInstall()=$ $\{rep[i] \gets rep[myLeader]\}$\label{VSm:adoptrep}\; 
			
			$followMcastRnd()$  \{$rep[i] \gets rep[myLeader]$\label{VSm:replicate}; $apply(state,rep[myLeader].msg)$; $input \gets \mathit{fetch}()$\label{VSm:fetchFol};\}
			
			\BlankLine
			
			\noindent \textbf{procedure} \label{VSm:procedure}
			$coordinateM\!castRnd()$ \textbf{do} \Begin{
				$apply(state, msg)$;
				$input\gets \mathit{fetch}()$\; \label{VSm:fetchCrd}
				
				\lForEach{$p_j\in C$}{\textbf{{if} $p_j\in view.set$~\textbf{then} $msg[j]\gets rep[j].input$ \textbf{else} $msg[j]\gets \bot$\label{VSm:collect}}}
				\textbf{write}$(\mathit{data},(i,\mathit{lLeader},rnd,rep[i].state))$\label{ln:writedata}\; 
				$rnd \gets rnd+1$\label{VSm:rndIncr};
				{\lIf{$rnd =\mathit{MAXINT}$}{$view.set\gets \bot$ \texttt{/* Forces a view change in line~\ref{VS:newviewsl}/}}}
			}
			
			\BlankLine
			
			\noindent \textbf{ do forever}  \texttt{/* use predefined periodicity */}  \Begin{
				$FD \gets failureDetector()$\label{VS:FDread}\; 
				\textbf{let} $(lDevices,lCloudlets, \mathit{lLeader},lGuards) := \mathbf{read}(\mathit{info})$\;
				
				
				\lIf{$\mathit{lLeader}.id = i \land  myLeader \neq i$\label{VS:newleader} }{$(status$, $propV$, $myLeader) \gets ( \textsf{Propose}, \langle(\mathit{lLeader},cnt=0)$, $FD$ $\cap$ $(lGuards$ $\cup\{ i \}) \rangle, i)$\label{VS:propose}}
				
				
				\lIf{$\mathit{lLeader}.id = i \land  myLeader = i \land ((status= \textsf{Multicast} \land view.set \ne  S)) \lor (status\neq \textsf{Multicast} \land propV.set \ne S)))$\label{VS:newviewsl}}{$(status, propV, myLeader) \gets ( \textsf{Propose}$, $\langle(\mathit{lLeader},$ $cnt$++$), S) \rangle, i)$, \textbf{where} $S:=FD\cap (lGuards\cup\{i\}$\label{VS:newview}}

				\lIf{$k \ne i \land i\in lGuards  \land k \in FD,$ \textbf{\emph{where}} $k=\mathit{lLeader}.id$}{$(myLeader,status)\gets  (k,rep[k].status)$}

				\If{$\mathit{lLeader}.id = i\land roundProceedReady()$\label{VS:crdConditions}}{
					\lIf{$status = \textsf{Multicast}$}{$coordinateM\!castRnd()$\label{VS:crdMulticast}}
					\lElseIf{$status = \textsf{Install}$}{$coordinateInstall()$\label{VS:crdInstall}}
					\lElseIf{$status = \textsf{Propose}$}{$coordinatePropose()$\label{VS:crdPropose}}
				}
				\ElseIf{$ \mathit{lLeader}.id \neq i \land i\in lGuards \land \mathit{lLeader}.id \in FD \land roundReadyToFollow()\label{VS:flrConditions}$}{
					\lIf{$status = \textsf{Multicast}$}{$followMcastRnd()$\label{VS:flrMulticast}}
					\lElseIf{$status = \textsf{Install}$}{$followInstall()$\label{VS:flrInstall}}
					\lElseIf{$status = \textsf{Propose}$}{$followPropose()$\label{VS:flrPropose}}
				}
				
				\lIf{$\mathit{lLeader}.id\neq i \land i\in lGuards \land \mathit{lLeader}.id \notin FD$\label{ln:leaderfail}}{
					$myLeader\gets \bot$;~\textbf{write}$(\mathit{data},(i,\mathit{lLeader},rnd,rep[i].state))$\label{ln:guardwrite}} 
				\lElseIf{$myLeader\neq \bot$}{$\mathbf{send}~rep[i]$ to $myLeader$}
				
				\lIf{$\mathit{lLeader}.id=i$}{$\forall k \in lGuards\cap FD$ \textbf{send} $\langle rep[i] \rangle$ \textbf{to} $p_k$\label{VS:send}}
				
				\lIf{$cnt =\mathit{MAXINT}$}{$\mathbf{send}(cloudID, \langle \textsf{RESET} \rangle)$\label{ln:cntMaxed}}
				
			}
			
			\noindent \textbf{ upon message} $m$ \textbf{arrival from} $p_j$ \textbf{do} $rep[j]\gets m$\; \label{VS:receive}
			
		\end{\algSizeSmall}
	\end{algorithm}
	

	\section{Correctness Proof}
	%
	%
	\label{sec:proof}
	{Our analysis demonstrates a constant time recovery from arbitrary transient faults. It} considers the \emph{interleaving model}~\cite{DBLP:books/mit/Dolev2000}, in which the node's program is a sequence of \emph{(atomic) steps}. Each step starts with an internal computation and finishes with a single communication operation, \ie message $send$ or $receive$. The \emph{state}, $s_i$, of node $p_i \in \sP$ includes all of $p_i$'s variables as well as the set of all incoming communication channels. Note that $p_i$'s step can change $s_i$ as well as remove a message from $channel_{j,i}$ (upon message arrival) or add a message in $channel_{i,j}$ (when a message is sent). The term \emph{system state} refers to a tuple of the form $c = (s_1, s_2, \cdots,  s_n)$ (system configuration), where each $s_i$ is $p_i$'s state (including messages in transit to $p_i$). An \emph{execution (or run)} $R={c_0,a_0,c_1,a_1,\ldots}$ is an alternating sequence of system states $c_x$ and steps $a_x$, such that each $c_{x+1}$, except $c_0$, is obtained from the preceding one, $c_x$, by the execution of step $a_x$. We say that execution $R$ is legal if it satisfies the task specifications throughout $R$. We say that a system state $c$ is safe if every execution that start from $c$ is legal. Definition~\ref{def:safe} considers a system state that Theorem~\ref{thm:recovery} shows to be safe.  
	
	\begin{definition}[Safe system state]
		\label{def:safe}
		We say that the system state $c$ is safe if the following hold.
		%
		(1) Let $p_i \in C$ and $p_j \in S$, such that $(j, m_j) \in myIoT(devices_{cloudID},cloudlets_{cloudID})$. It holds that	$cloudletList_j = cloudletList(j,cloudlets_{cloudID})\land(lastUpdate_j \leq clock_j())$. Moreover, $devices_{cloudID} =\{ (k,\bullet) \in deviceSet_i\}\land ((z,t,\bullet)\in \emph{\text{agreegateInfo}}_j \implies p_z \in C \land t \leq clock_i()) \land (j,\bullet,m_j) \in \emph{\text{agreegateInfo}}_j$.
		%
		%
		%
		%
		(2) The value of $\mathit{msgseq}_i$, $msgc_i$ and $msgtoiot_i$ is greater or equal to any value of $\mathit{msgseq}$, $msgc$, and respectively, $msgtoiot$ fields associated with $p_i$ in messages and cloudlets.
		%
		%
		(3) $|A|=1$, where $A=\{(v,su,r,sa,m): p_i,p_j \in C \land (v,su,r,sa,m,\bullet) = r \in \{rep_{i}[j],rep_{i,j}\} )\}$, such that $rep_{i,j}$ is a message that was sent in line~\ref{VS:send} from $p_i$ to $p_j$. Moreover, $msg_{leader_{cloudID}.id}[k]=input_k$, where $p_k \in leader_{cloudID} \cup guards_{cloudID}$. 		
		(4) No counter has reached $\text{MAXINT}$ and there are no $\langle \textsf{RESET}\rangle$ messages. 
		%
		%
	\end{definition}

	We say that an execution is {\em fair} if every step that is applicable infinitely often is executed infinitely often. Theorem~\ref{thm:recovery} demonstrates the required properties for self-stabilization and use the term \emph{(asynchronous) cycles} of a fair execution $R$. A cycle is the shortest prefix of $R$ in which every non-failing node $p_i$ performs a completed iteration of node $p_i$'s do forever loop, all messages that $p_i$ sent during that iteration were delivered, and all of the iteration's requests were replied. 
	
	\begin{theorem}
		\label{thm:recovery}
		The system's state is safe within $\bigO(1)$ cycles. 
	\end{theorem}
	
	\begin{proofsketch}
		The proof considers the predicate {$\textit{pred}=\{\bot \} \neq ( \{\mathit{info} \} \cup \{ \mathit{infoAck}[k] : k \in C\setminus suspectedCloudlet_{cloudID}(C) \})$}. We start by considering an execution in which $\textit{pred}$ holds throughout $R$, and thus $p_{cloudID}$ does not executes lines~\ref{ln:empWrite} and~\ref{ln:botWrite}. Under this assumption, we show that items~1 to~4 of Definition~\ref{def:safe} hold within $\bigO(1)$ cycles. As a completely case, we consider a starting system state in which $\textit{pred}$ does not hold, and show that, within $\bigO(1)$ cycles, the $\textit{pred}$ holds.
		
		\smallskip \noindent \textbf{Item 1.}
		Let $(p_i,p_j) \in C \times S$. Within $\bigO(1)$ cycles, the cloud $p_{cloudID}$ updates the $devices$ and $cloudlets$ fields in $\mathit{info}$ (line~\ref{ln:writeInfo}). Within $\bigO(1)$  cycles, $p_i$ reads $devices$ and $cloudlets$ (line~\ref{ln:readInfo}) and send $\langle \bullet, cloudletList(j,lCloudlets), m \rangle$ to IoT $p_j$ (line~\ref{ln:msgtoiotSend}), such that $(j, m_j) \in myIoT(devices_{cloudID},cloudlets_{cloudID})$. When that message arrives, $p_j$ stores it in $cloudletList_j$ and $\mathit{cloudletModel}_j$ as well as updates $lastUpdate_j$ with the arrival time (line~\ref{ln:iot:arrival}). Thus, $cloudletList_i$ $=$ $cloudletList(i,cloudlets_{cloudID})$ $\land$ $(lastUpdate_i$ $\leq$ $clock_i())$. Lines~\ref{ln:deviceSetUpdate} and~\ref{ln:deviceSetArrival} implies $devices_{cloudID} =\{ (k,\bullet) \in deviceSet_j\}$ and line~\ref{ln:agre} implies $((z,t,\bullet)\in \emph{\text{agreegateInfo}}_j \implies p_z \in C \land t \leq clock_i()) \land (j,\bullet,m_j) \in \emph{\text{agreegateInfo}}_j$.
		
		%
		
		\smallskip \noindent \textbf{Item 2.~~} 
		Suppose that in $R$'s starting state, Item~3 does not hold with respect to a $p_i$'s field. Within $\bigO(1)$ cycles, any message containing $\mathit{msgseq}$, $msgtoiot$ or $msgc$ arrive to its destination $p_j$. Thus, for the sake of a simple presentation, we focus on the case in which Item~3 does not hold in node $p_j$ with respect to a field that is associated with node $p_i$. We observe that within $\bigO(1)$ cycles, $p_i$ and $p_j$ complete a message round-trip that include this filed. In detail, these message are sent in lines~\ref{ln:msgseqSend},~\ref{ln:msgtoiotSend} and~\ref{ln:msgcSend} and received in lines~\ref{ln:msgseqRec},~\ref{ln:msgtoiotRec} and~\ref{ln:msgcRec}, respectively. Note that whenever $p_i$ receives any such message, $p_i$ updates the local value with the received one, in case the latter is greater than the former.      
		
		\smallskip \noindent \textbf{Item 3.~~}
		%
		%
		%
		Within $\bigO(1)$  cycles, $leader$ and $guards$ are set by $p_{cloudID}$ (line~\ref{ln:writeInfo}) and all cloudlets read these values (line~\ref{ln:readInfo}). {We show that if a new leader has been put in place (or the view has become inconsistent), the leader installs a new view which includes itself and the guards. For this purpose, it first proposes this view (line~\ref{VS:crdPropose}), which it is accepted by the guards (line~\ref{VS:flrPropose}) within $\bigO(1)$ cycles. Then, within $\bigO(1)$ cycles it updates $rep[]$ and installs this view (lines~\ref{VS:crdInstall} and \ref{VS:flrInstall}), in which the guards have updated their $rep[]$s based on the one of the leader.} After that the leader resumes the round-base updates for maintaining the state among itself and the guards {(lines~\ref{VS:crdMulticast}, \ref{VS:flrMulticast}, and \ref{VSm:procedure}--\ref{VSm:rndIncr}), hence stabilizing the state machine replication}. Moreover, it aggregates $input_k:p_k \in leader_{cloudID} \cup guards_{cloudID}$, such as $msg_{leader_{cloudID}.id}[k]=input_k$.
		
		\smallskip \noindent \textbf{Item 4.~~} 
		Suppose that in $R$'s starting state, Item~3 does not hold at node $p_i$. We observe that within $\bigO(1)$ cycles, either Item~3 holds and Item~4 does not hold, or both hold. Moreover, within $\bigO(1)$ additional cycles, $\langle \textsf{RESET}\rangle$ arrives to $p_{cloudID}$ and the assumption above does not hold (line~\ref{ln:reset}). Thus, the rest of the proof considers complementary case.
		
		\medskip
		
		For the case that is competently to the assumption that appears in the proof start, suppose, towards a contradiction, that $\textit{pred}$ does not hold in $R$'s starting state. Moreover, suppose that any prefix $R'$ of $R=R'\circ R''$ that has $\bigO(1)$ cycles, does not have a matching suffix $R''$ during which the predicate $pred$ holds. Since $\mathit{pred}$ does not hold during $R'$, $p_{cloudID}$ does not executes lines~\ref{ln:newIotEmpty} to~\ref{ln:writeInfo} during $R'$. Therefore, it must executes either line~\ref{ln:empWrite} or~\ref{ln:botWrite} for a constant number of times during $R'$. 
		
		Suppose that $p_{cloudID}$ does not execute line~\ref{ln:empWrite} during $R'$. Thus, within $\bigO(1)$ cycles, $p_{cloudID}$ executes repeatedly line~\ref{ln:botWrite} until the if-statement condition of line~\ref{ln:empWrite} holds. Then, the if-statement condition of line~\ref{ln:botWrite} does not hold again during $R$, and, within $\bigO(1)$ cycles, the if-statement condition of line~\ref{ln:botlInfot} holds. Thus, the system reaches $R''$ within $\bigO(1)$ cycles.~\end{proofsketch}

	\section{Evaluation}
	\label{sec:evaluation}

	The previous section details the correctness proof of our self-stabilizing algorithmic process which, in contrast to the current state-of-the-art, shows that \textit{even in the presence of failures a fog ecosystem can always recover in constant time and compute analytic insights from IoT data}. 
	
	This section introduces a comprehensive evaluation of the effectiveness and runtime overhead of our framework. First, we measure \textit{information delay} that is the time interval required for IoT data to be propagated in the network for analytics to be correctly derived in the presence of multiple and different failures (\eg cloudlet fail-stop, communication link drops). Second, we measure the additional \textit{runtime footprint} that our framework incurs to exemplary state-of-the-art distributed control planes (\eg istio). This provides a detailed overview of what is the cost, in terms of network overhead, of maintaining \textit{data freshness} and analytic computation \textit{correctness} in the presence of failures.
	\textit{Results show that with our self-stabilizing framework, control planes are able to compute analytics correctly with the information delay maintained relatively stable despite of the presence of failures, while the network overhead scales linearly towards the IoT load, as required by O1-O4 (Section~\ref{sec:system})}.
	
	\begin{figure}[t]
		\centering
		\includegraphics[width=0.81\linewidth]{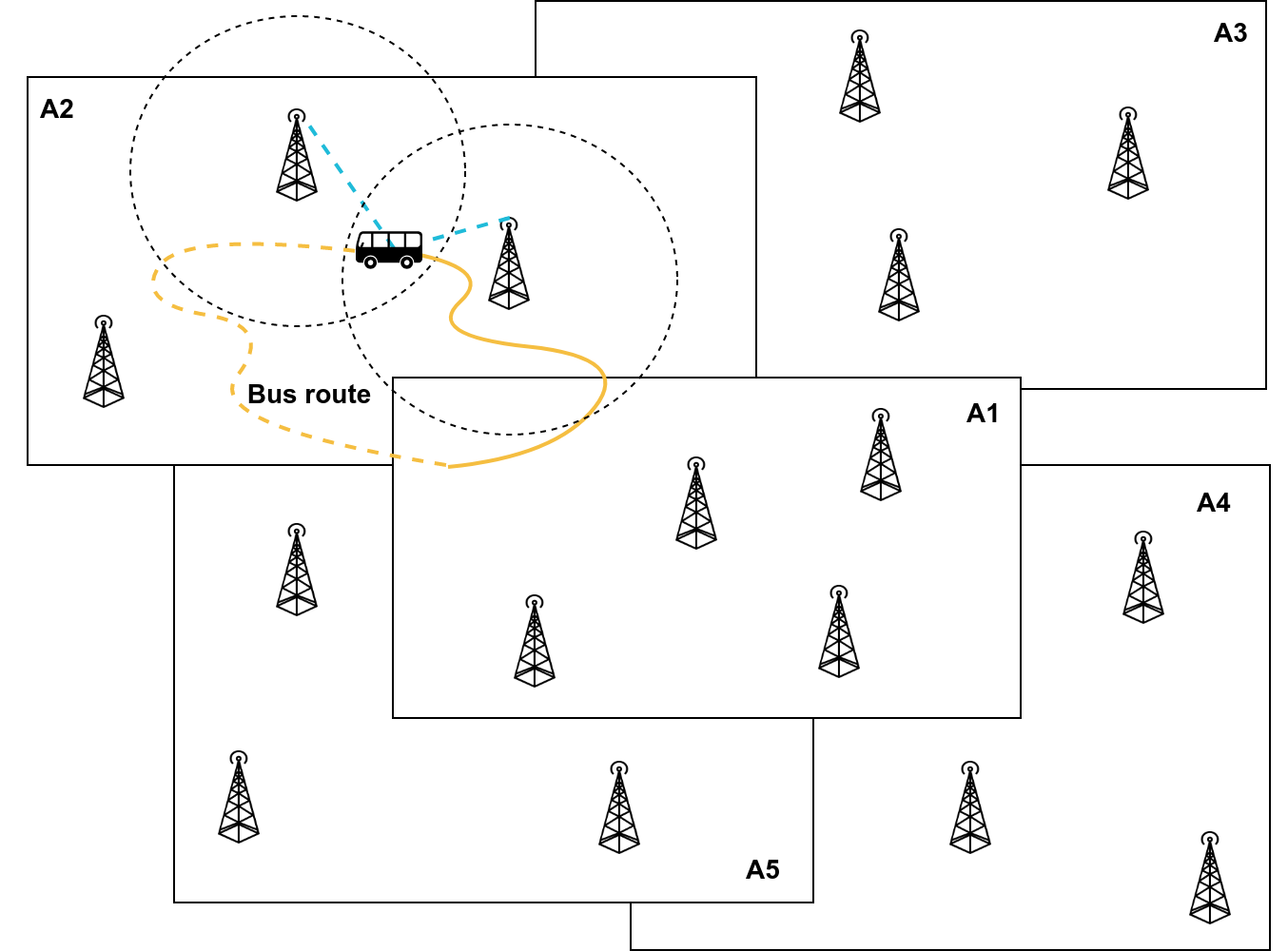}
		\caption{High-Level Overview of the Bus Network Topology}
		\label{fig:extopo}
	\end{figure}

	For the experimentation, we introduce a real-world use-case of a smart city Bus Network Service (BNS) evaluated under various execution scenarios. We opt to focus on experiments that use a publicly available and real-world workload to truly reveal the strengths of our framework and its ability to deal with high workload.
	Specifically, the workload originates from the Dublin smart city Bus Network Service~\cite{dublinbuses}, comprised of 40GB of compressed data, tracking for 1 month the bus routes of 968 buses (Jan. 2013). Each bus is equipped with a GPS tracking device recording every $1s$ location coordinates and the current bus route delay. Figure~\ref{fig:extopo} depicts a high-level overview of the BNS topology, where 16 cloudlets are deployed across Dublin's major city regions, denoted  for clarity as $A_x$, to decentralize the BNS and increase the system responsiveness. 
	We note that a bus route may span across different city regions and a bus can be connected to multiple cloudlets depending on the cloudlet coverage. Each cloudlet serves as an analytics engine that aggregates local bus updates and propagates an alert to traffic operators (central cloud service) when 10 or more buses in a city area are reporting, in a 5min sliding window, delays over one standard deviation from the previous weekly mean.
	
	
	
	To 
	experiment with large-scale deployments and ensure both result reproducability and algorithm adoption, we have designed a simulation testbed inspired by Kompics
	~\cite{Arad2012}, an open-source distributed systems message-passing component model, and extended the entity behavior model to facilitate fault models over a control plane in fog computing ecosystems\footnote{~Coded artifacts, testbed configurations and workload (including dataset and faults) are available here: \url{https://github.com/UCY-LINC-LAB/Self-Stabilization-Edge-Simulator}}. The testbed is run in an Openstack private cloud on a server configured with 16VCPU clocked at 2.66GHz, 16GB RAM and 260GB disk. The network configuration between testbed entities adopts a gaussian kernel with the following mean values: (i) Cloudlet-to-Cloud latency 100ms; (ii) IoT-to-Cloud latency 250ms; (iii) intra-region Cloudlet-to-Cloudlet latency 10ms; (iv) inter-region Cloudlet-to-Cloudlet latency 100ms; and (v) IoT-to-Cloudlet latency 20ms.
	%
	%
	We opt for these specific capabilities so that the testbed resembles an actual geo-distributed fog deployment over a city environment. All simulation scenarios are run 100 times with cloudlets and IoT devices starting at randomized time intervals. For the IoT device placement, we have implemented the registration interface of Algorithm 2 so that when an IoT device (\eg a bus) requests to join the network, the central authority (\eg the cloud) responds with a list of valid cloudlets that are the ``closest'' to the device in the device's operating (city) region. The same strategy will hold for when the device has changed it's operating region (\eg bus moves from $A_1$ to $A_2$). Finally, the selection of the leader and the guards was done randomly, since our  cloudlets are homogeneous.
	
	For the widespread experimentation of different fault scenarios over the testbed, we adopt the Netflix Chaos Monkey framework~\cite{chaosmonkey}. This enables the configuration and (random) selection of faults and entities to infest at given time intervals, or at random, depending on the evaluation scenario. Unless otherwise stated, the aforementioned topology and network configuration will be considered as the {\bf\em baseline configuration}. 

	\subsection{Information Delay}
	\label{sec:info-delay}
	In this set of experiments, we show the effect of different failures to the timeliness of analytic computation. We consider four experiment runs with faults injected at random and examine how information delay is affected by: 
	\begin{itemize}[-]
		\item randomly failing a different number of regular cloudlets;
		\item failing the guards; 
		\item failing the leader;
		\item randomly dropping the communication link between IoT devices and cloudlets.
	\end{itemize}
	
	
	\begin{figure}[t]
		\centering
		\includegraphics[width=0.95\linewidth]{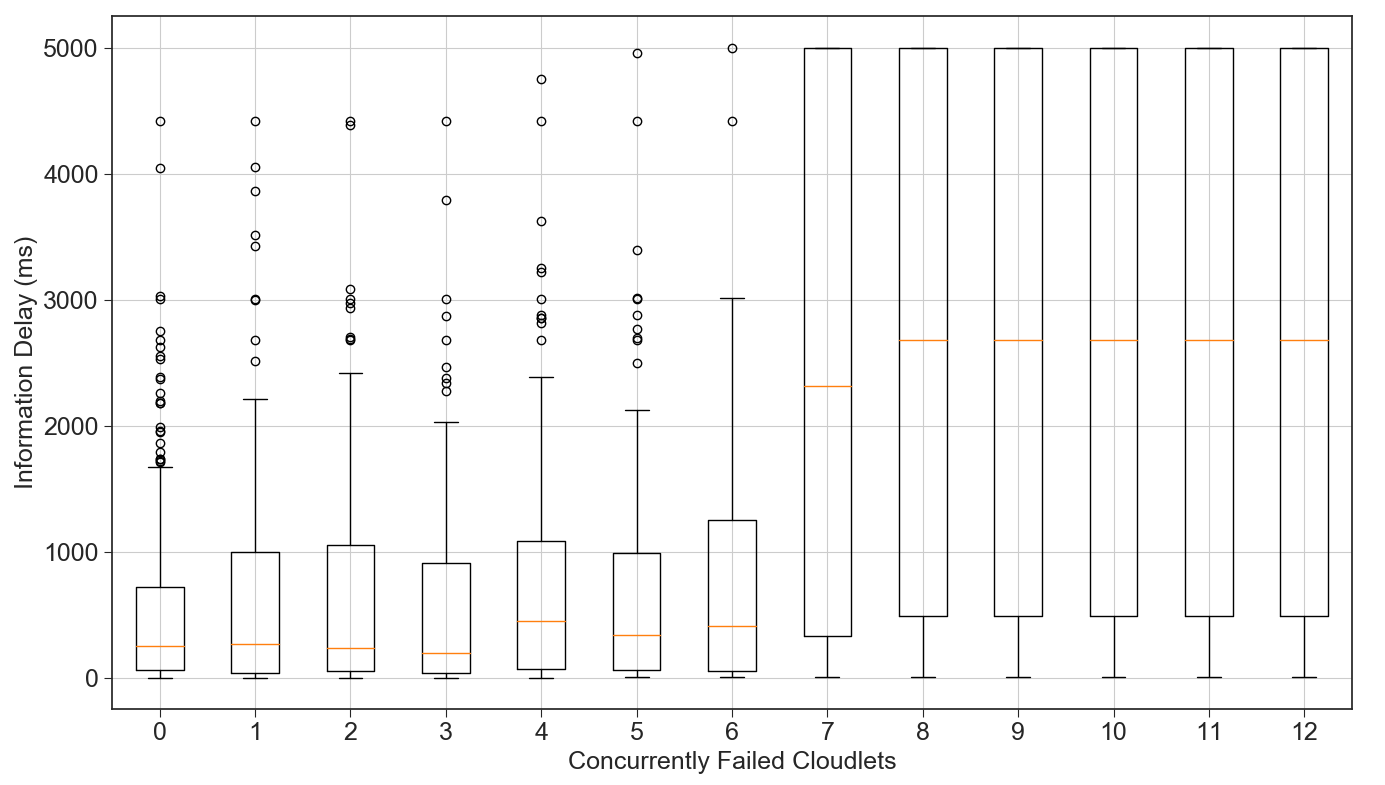}
		\caption{Information delay vs number of concurrent cloudlet failures}
		\label{fig:info-delay-cloudlet}
	\end{figure}
	
	
	Figure~\ref{fig:info-delay-cloudlet} depicts the information delay as the number of concurrently failing cloudlets increases. In this box-plot the median information delay is denoted by the line in the box, while the box length extends between the first and third quantile with outliers depicted as independent points. With zero cloudlets we denote the information delay in normal operation (without failures). From Figure~\ref{fig:info-delay-cloudlet}, we observe that information delay is not affected, despite slight deviations, while the number of failing cloudlets remains under 7. After this, randomly selecting concurrent cloudlets hinders the extreme case of wiping out all cloudlets of a city region. This results in added delay as IoT data for the specific region must be directly propagated to the cloud. For this experiment run, system recovery is only required when an IoT device is left with no cloudlet in its coverage. In this extreme case, the IoT device must contact the cloud to validate the registration. However, the involvement of the cloud naturally hinders a communication overhead. Thus, \textit{despite information delays for extreme cases of concurrent cloudlet failures, analytic computation is correct at all times while the system recovers from faults in a bounded number of communication rounds, as required by O1 and O3 (Section~\ref{sec:system})}. 
	
	\begin{figure}[t]
		\centering
		\includegraphics[width=0.99\linewidth]{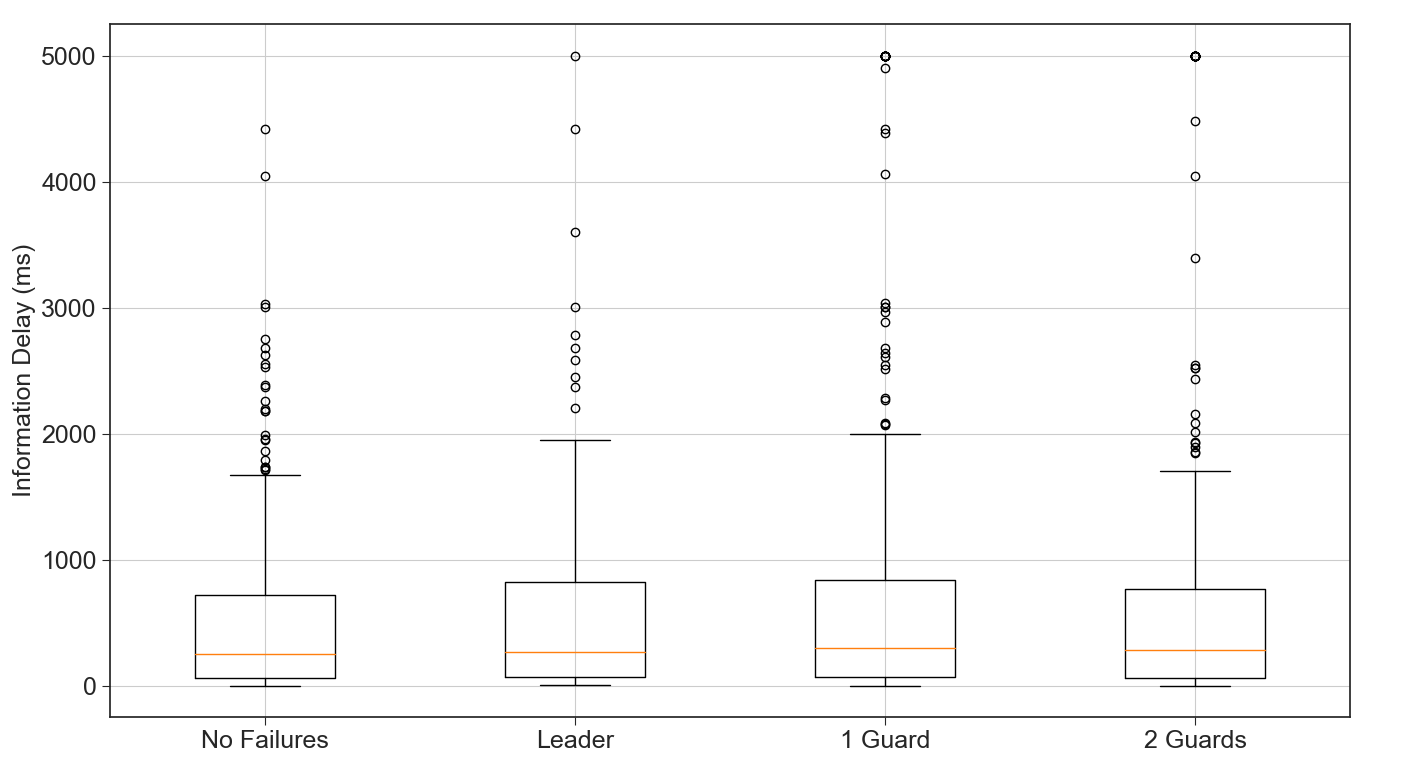}
		\caption{Information delay vs number of  concurrent guard and leader failures} 
		\label{fig:info-delay-guards}
	\end{figure}
	
	Figure~\ref{fig:info-delay-guards} depicts how information delay is affected by the failure of the control plane guards and leader when the baseline deployment is configured with two guards. We observe that the timeliness of analytics computation is neither affected by the failure of the leader or the guards. This  concurs with the correctness proof that shows that, \textit{the self-stabilizing fog ecosystem can return back to a legal state within $\bigO(1)$ time, which is sufficient to propagate information without delay, as required by O4 (Section~\ref{sec:system})}.
	

	\begin{figure}[t]
		\centering
		\includegraphics[width=0.98\linewidth]{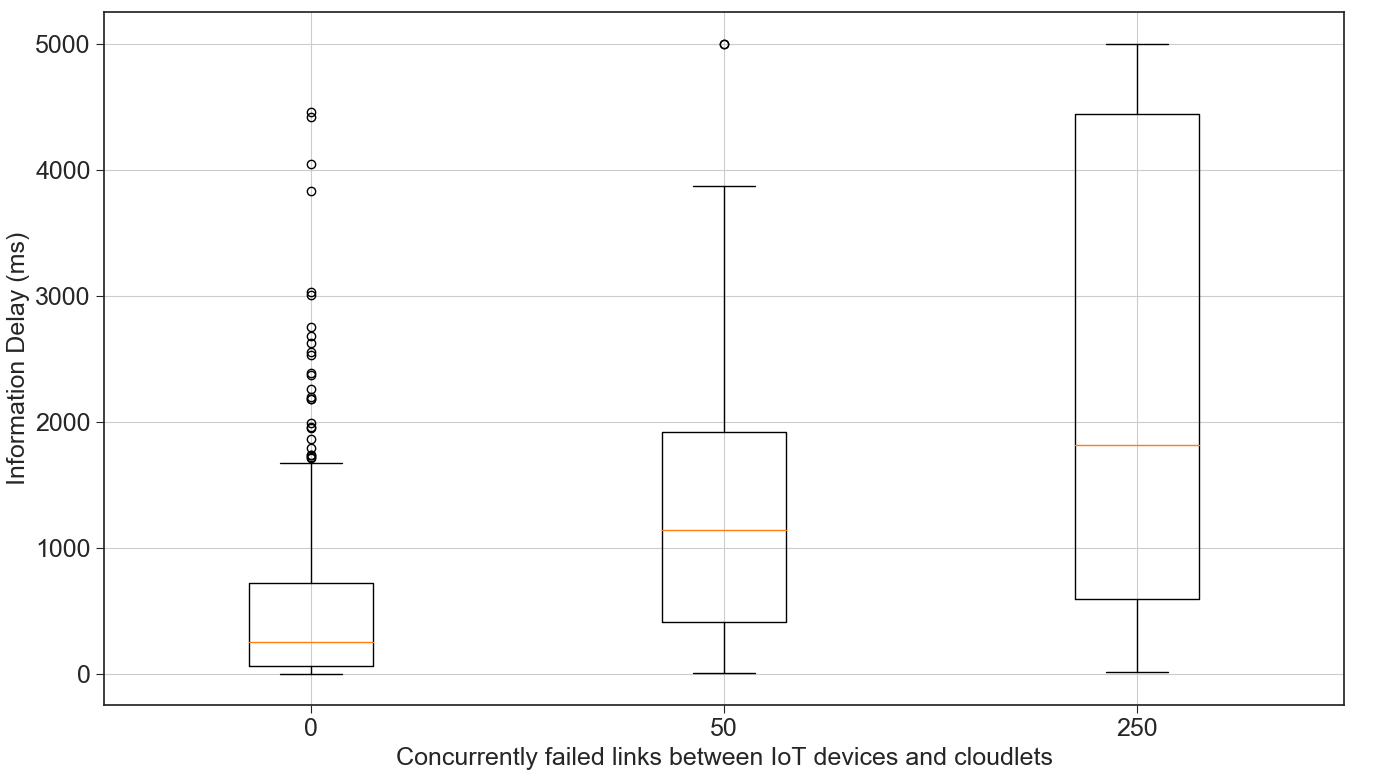}
		\caption{Information delay vs concurrent fail-stop IoT-Cloudlet network links} 
		\label{fig:info-delay-commlink}
	\end{figure}
	
	The next experiment run studies how information delay is affected by the temporary drop of the network link between IoT devices and cloudlets. To achieve this, we artificially block for a predefined interval 
	the link between affected IoTs and cloudlets in each region, thus maintaining only the link with the cloud.
	In Figure~\ref{fig:info-delay-commlink} we observe that the information delay increases as more devices experience a link drop. This occurs because the affected IoTs detect the link absence and, thus, must communicate with the cloud for updates which takes more time. Still, \textit{analytics are computed without corrupted or missing IoT data}. This extreme case, of failing all the communication links among IoT and cloudlets, highlights the importance of having a sufficient amount of cloudlets in each region to cope with concurrent link failures.
	
	\begin{figure}[t]
		\centering
		\includegraphics[width=1\linewidth]{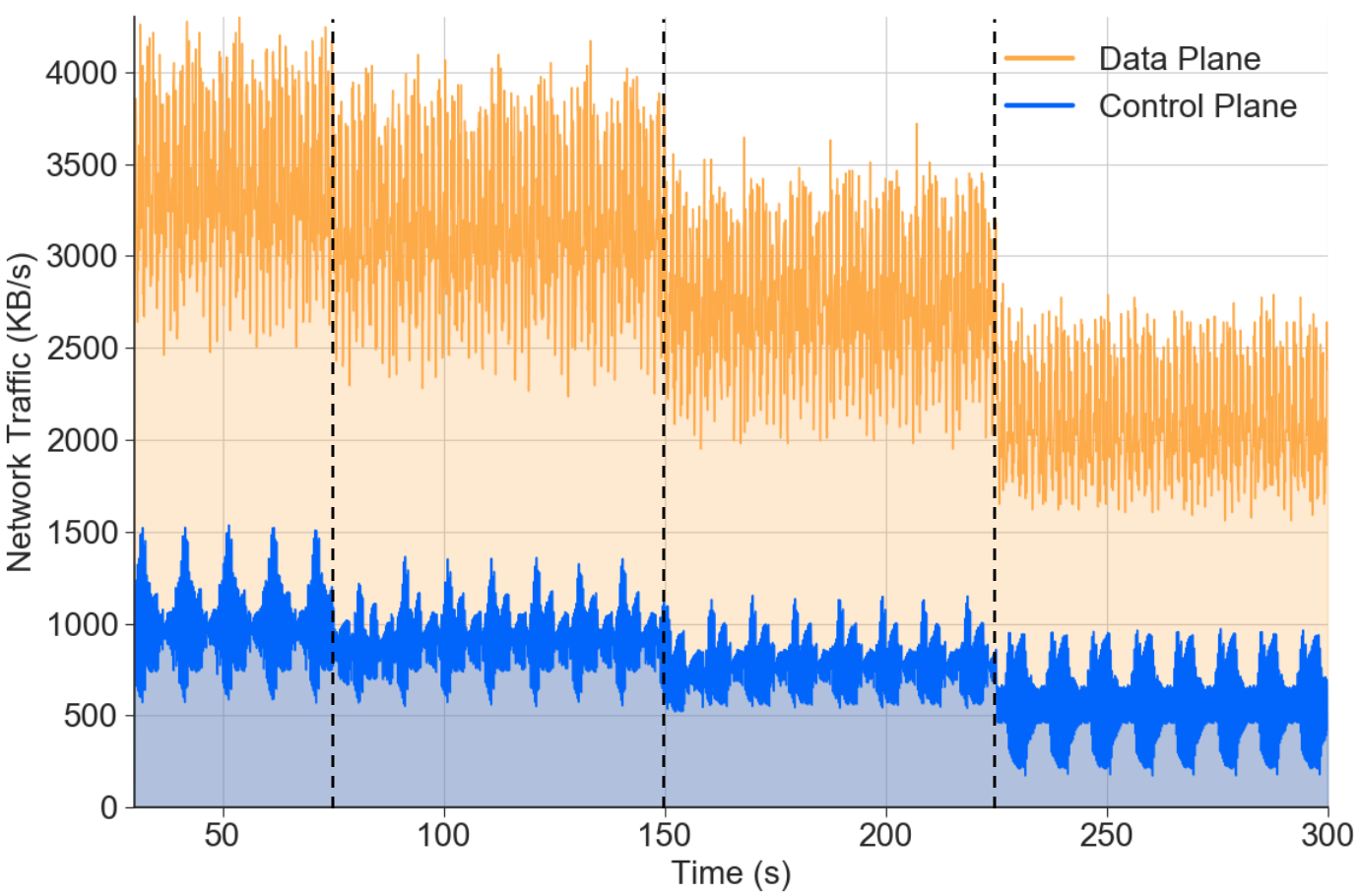}
		\caption{Control and Data Plane Traffic} 
		\label{fig:net-overhead}
	\end{figure}
	
	\subsection{Runtime Footprint}
	\label{sec:runtime-footprint}
	In this set of experiments, we provide an analysis depicting the network overhead of different components comprising our framework and the experiment testbed. Figure~\ref{fig:net-overhead} depicts the network traffic over the data and control plane for a simulation run of the baseline configuration when random failures of the cloudlets' leader, guards, and cloudlets are introduced. 
	The figure depicts the network overhead for 5min where the $30s$ bootstrap period is omitted. First, we observe four distinct segments (separated by vertical lines). During each segment our framework maintains a stable message exchange rate for both planes, with the data plane traffic approximately x3.5 higher than the control plane traffic. In the first segment ($30s$ to $75s$) the system exhibits no faults. At the 75th second, the leader fails and we observe a slight drop in both the control plane traffic (from 950KB/s to 850KB/s) and the data plane (from 3300KB/s to 3100KB/s). When the cloud discovers the leader failure, it elects a new leader at the 88th second and the system recovers back to a legal state, with a slight increase of the control plane traffic (900KB/s). Next, at the 150th second the two guards fail and the control plane traffic falls to 700KB/s while the data plane traffic falls to 2600KB/s. As before, the cloud elects two new guards and the control plane traffic stabilizes at 750KB/s. Finally, at the 225th second (4th segment) three cloudlets fail and both control and data traffic drop to 550KB/s and 2100KB/s, respectively. These results show that a constant number of messages is exchanged, validating the objectives O1 and O3 (Section~\ref{sec:system}).

	Next, we show that the control and data plane network traffic scales linearly towards the number of different system entities, as required by O2. Table \ref{tbl:traffic} shows the results of different configurations
	in percentage increments from the baseline.
	
	\noindent\textbf{Guards}.
	We observe that the overhead of adding guards increases linearly. Specifically, each additional guard adds an overhead in the range of $4.75-5.68\%$ for the control plane traffic and $4.82-5.02\%$ for the data plane traffic. It is worth pointing out that the previous experiment in Figure~\ref{fig:info-delay-guards} showed that even with all the guards failing concurrently, the information latency remains stable, and therefore, for the studied baseline configuration, having two guards balances well the trade-off between overhead and information delay.
	
	\noindent\textbf{Cloudlets}.
	The overhead of adding extra cloudlets, for redundancy purposes, scales linearly while the IoT load remains stable. Specifically, each additional cloudlet adds an overhead in the range of $7.96-9.03\%$ for the control plane, and for the data plane the increment is approximately $6.2\%$. Obviously, the trade-off is straightforward. Increasing the cloudlets, decreases the probability of delaying information propagation for a city region, \eg as in the case of Figure~\ref{fig:info-delay-cloudlet} after 7 cloudlets, at the cost of higher network traffic.

	\noindent\textbf{IoTs}.
	By increasing the workload (IoT devices), again, the network overhead is linearly increased. Each additional IoT device adds a $0.094\%$ overhead on the control plane traffic, while for the data plane the increment ranges between $0.085-0.087\%$. This increase is attributed to the fact that each cloudlet communicates with more IoT devices.
	
	\begin{table}[]
		\centering
		\begin{tabular}{@{}lcc@{}}
			\toprule
			\begin{tabular}[c]{@{}l@{}}System\\ Change\end{tabular} & \begin{tabular}[c]{@{}l@{}}Control Plane Traffic \\ Change Compared \\ to Baseline (\%)\end{tabular}  & \begin{tabular}[c]{@{}l@{}}Data Plane Traffic \\ Change Compared \\ to Baseline (\%)\end{tabular} \\ \midrule
			3 Guards & 4.75 & 5.02 \\
			4 Guards & 10.38 & 9.64 \\
			5 Guards & 17.04 & 14.50 \\
			6 Guards & 22.06 & 19.44 \\ \midrule
			20 Cloudlets & 31.85  & 25.03  \\
			25 Cloudlets & 76.63 &  56.38  \\
			30 Cloudlets & 126.48 & 87.85 \\ \midrule
			1500 IoT & 46.76  & 43.25 \\
			2000 IoT &  93.59  & 85.58 \\
			2500 IoT & 140.18 &  127.51 \\ \midrule
		\end{tabular}
		\caption{Network Traffic Overhead over Topology Changes }
		\label{tbl:traffic}
	\end{table}

	\section{Conclusions}
	In this paper we introduced a fault-tolerant  framework  for distributed control   planes   that   enables   fog   services   to cope with  a very broad fault model. To this end, we presented self-stabilizing algorithms that guarantee automatic recovery within a constant number of communication rounds without the need for external (human) intervention. Using real-world data and actual queries of interest from an intelligent transportation service, we demonstrate the performance gains of our framework, and thus the promise of self-stabilization in fog computing. 
	Our results show that despite  information  delays for  extreme  cases  of  concurrent  cloudlet  failures,  analytic computation is correct, while the network  overhead  is  proportional  to  the number of cloudlets, guards, and devices. 
	We believe that our self-stabilizing framework is applicable to a wide range of fog services requiring strong fault-tolerance guarantees.
	

	\section*{Acknowledgement} This work is partially supported by the EU Commission through RAINBOW  871403 (ICT-15-2019-2020) project and by VINNOVA, the Swedish Government Agency for Innovation Systems, project ``\href{https://www.vinnova.se/p/autospada-automotive-stream-processing-and-distributed-analytics-oodida-phase-2/}{Automotive Stream Processing and Distributed Analytics (AutoSPADA)}'' in the funding program FFI: Strategic Vehicle Research and Innovation (DNR 2019-05884).


\end{document}